\documentclass[manuscript]{acmart}

\usepackage{color}
\usepackage{tikz}
\usepackage{soul}
\usepackage{amsmath}
\usepackage{tabu}

\newcommand{\Cross}{\mathbin{\tikz [x=1.4ex,y=1.4ex,line width=.2ex] \draw (0,0) -- (1,1) (0,1) -- (1,0);}}

\newcommand{\name}[1]{Infographics Wizard}

\AtBeginDocument{%
  \providecommand\BibTeX{{%
    \normalfont B\kern-0.5em{\scshape i\kern-0.25em b}\kern-0.8em\TeX}}}

\setcopyright{acmcopyright}
\copyrightyear{2021}
\acmYear{2021}
\acmDOI{10.1145/1122445.1122456}

\acmJournal{TOG}
\acmVolume{37}
\acmNumber{4}
\acmArticle{111}
\acmMonth{8}

\acmSubmissionID{6197}

\begin{document}

\title{User centric Semi-Automated Infographics Authoring and Recommendation}


\author{Anjul Tyagi}
\affiliation{%
  \institution{State University of New York}
  \city{Stony Brook}
  \state {New York}
  \country{USA}}
\email{aktyagi@cs.stonybrook.edu}

\author{Jian Zhao}
\affiliation{%
  \institution{University of Waterloo}
  \city{Waterloo}
  \state {Ontario}
  \country{Canada}}
\email{jianzhao@uwaterloo.ca}

\author{Pushkar Patel}
\affiliation{%
  \institution{Indian Institute of Information Technology}
  \city{Vadodara}
  \state {Gujarat}
  \country{India}}
\email{201851094@iiitvadodara.ac.in}

\author{Swasti Khurana}
\affiliation{%
  \institution{Indian Institute of Information Technology}
  \city{Vadodara}
  \state {Gujarat}
  \country{India}}
\email{201852020@iiitvadodara.ac.in}

\author{Klaus Mueller}
\affiliation{%
  \institution{State University of New York}
  \city{Stony Brook}
  \state {New York}
  \country{USA}}
\email{muelleri@cs.stonybrook.edu}








\begin{abstract}
Designing infographics can be a tedious process for non-experts and time-consuming even for professional designers. 
Based on the literature and a formative study, we propose a flexible framework for automated and semi-automated infographics design.
This framework captures the main design components in infographics and streamlines the generation workflow into three steps, allowing users to control and optimize each aspect independently.
Based on the framework, we also propose an interactive tool, \name{}, for assisting novice designers with creating high-quality infographics from an input in a markdown format by offering recommendations of different design components of infographics. 
Simultaneously, more experienced designers can provide custom designs and layout ideas to the tool using a canvas to control the automated generation process partially. 
As part of our work, we also contribute an individual visual group (VG) and connection designs dataset (in SVG), along with a 1k complete infographic image dataset with segmented VGs. This dataset plays a crucial role in diversifying the infographic designs created by our framework.
We evaluate our approach with a comparison against similar tools, a user study with novice and expert designers, and a case study. Results confirm that our framework and \name{} excel in creating customized infographics and exploring a large variety of designs.

\end{abstract}



\begin{teaserfigure}
    \includegraphics[width=\textwidth]{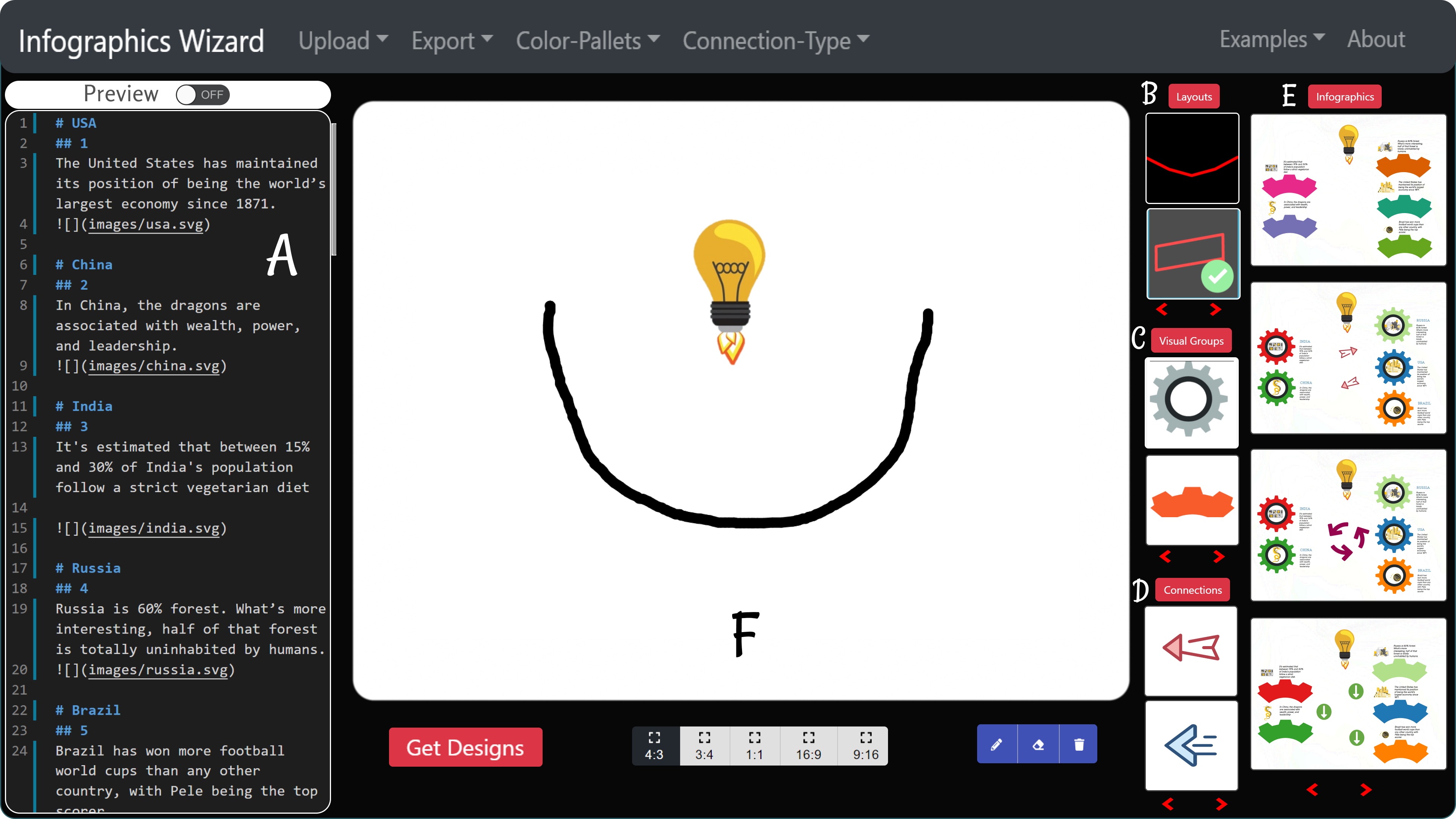}
    \caption{\name{} implements a flexible framework for full- and semi-automated infographic generation. Based on user's input in the markdown format (A), \name{} generates various recommendations for different main design components of infographics, including the Visual Information Flow layout (B), the design of individual Visual Groups (C), and the connecting elements between groups (D). The user can then explore these design alternatives (E) and assemble a final infographic of their desire. More experienced designers can optionally provide a main pivot graphic, the general information flow, or both on a canvas (F) via direct manipulation to control the generation and recommendation of infographic design.
 }
  \label{fig:teaser}
\end{teaserfigure}
\maketitle

\section{Introduction}\label{s:introduction}

Infographics have been widely used in various areas like fashion, advertisement, and business to convey complex data-driven narratives aesthetically by following well-studied design principles of human perception \cite{lankow2012infographics, dick2014interactive}. 
To convey information about a particular idea, designers generally divide infographics into separate repeated design components---\textit{Visual Groups (VGs)}---each carrying a specific piece of information \cite{lu2020exploring}. 
These design pieces are then organized together in a logical sequence---\textit{Visual Information Flow (VIF)}---which forms an infographic \cite{lu2020exploring}. They are further enhanced with design patterns according to eye movements, attention memorability, and aesthetics principles \cite{wang2018infonice, kumar2019task, niebaum2015infographics, borkin2015beyond,haroz2015isotype}. 
However, it is challenging for non-experts to develop compelling infographic designs because: (1) professional tools such as Adobe Illustrator have a steep learning curve, and (2) creating an infographic depends on many factors and requires sufficient expertise and experience.

To address these challenges, several fully-automated infographic design tools exist. For example, Text2Viz \cite{cui2019text} generates infographics from proportional-based text with several ``blueprint'' templates; DataShot \cite{wang2019datashot} only generates fact sheets from data with a tiled layout. While convenient, these tools are often based on a limited set of templates, lacking the variety of designs they can generate.  
Additionally, semi-automated tools allow the user to apply their design ideas with some cues and suggestions (e.g., \cite{brehmer2019timeline,chen2019towards,wang2018infonice,coelho2020infomages}). However, most of the proposed techniques focus on specific types of infographics or pieces of design. For example, Timeline Storyteller \cite{brehmer2019timeline} and Chen \textit{et al.}'s system \cite{chen2019towards} focus on timeline infographics; InfoNice \cite{wang2018infonice} just supports on data mark authoring; and ICONATE \cite{zhao2020iconate} assists with merely compound icon design.

In this work, we take a holistic view of infographic design by leveraging the concepts of VIF, and VG \cite{lu2020exploring}. 
We propose a flexible framework that decouples the general structure of infographics and offers support for novice designers by taking the benefits of design automation while providing the freedom to control major information pieces.
In specific, we discover four main design components in infographics, including (1) \textit{VIF layouts} that represent the backbone structure of the infographic, (2) \textit{VG designs} that determine repeating design components holding a specific piece of information, (3) \textit{pivot graphics} that set the stage of other design components and the overall infographic, and (4) \textit{connecting elements} that bind individual VGs together (or to the pivot graphic). 
Decoupling infographic design with such a framework allows each part to be optimized individually in fully- and semi-automated authoring tools.
This framework is the result of a formative study with 10 participants from a variety of backgrounds, through analyzing their perception of existing infographics and workflows of creating infographics. 
To foster future research in this direction, with the framework, we contribute an infographic dataset of 1K images with annotated VG designs, using Amazon Mechanical Turk \cite{mturk}. Along with this 1K segmented VGs in complete infographics images, we also release the extracted VGs in the SVG forms, maximizing the utility and benefits of the dataset. 

We realize this framework through developing a highly-interactive tool, called \textit{\name{}}, for rapid prototyping and design exploration of infographics (Figure~\ref{fig:teaser}). 
The framework indicates a three-stage pipeline that is implemented in \name{}, including: (1) recommending VIF layouts from given information, (2) suggesting various VG designs, and (3) generating connections between VGs to complete the infographic. 
Inspired by the prevalence of markdown languages in many applications such as web development and Jupyter Notebooks, \name{} allows the user to separate the manipulation of content and presentation in infographics design.   
In particular, the user specifies the content of each VG of the designed infographic in a markdown format (Figure~\ref{fig:teaser}A), without worrying about the layout or appearance.
Then, \name{} fulfills the three-stage pipeline by recommending appropriate VIF layouts (Figure~\ref{fig:teaser}B), VG designs (Figure~\ref{fig:teaser}C), and connecting elements (Figure~\ref{fig:teaser}D). Thus, the user can rapidly explore different design alternatives of each design component and obtain various final infographics flexibly and easily (Figure~\ref{fig:teaser}E).
While we develop these three steps to be completely automatic based on the input, designers have much freedom to intervene in the infographics generation and recommendation process. For example, the user can choose to upload a customized pivot graphic, a custom VG design, and optionally sketch a rough information flow for placing the VGs (Figure~\ref{fig:teaser}F).



We validated \name{} through a comprehensive evaluation containing three parts. First, we compared \name{} with various existing tools based on the Analysis of Competing Hypotheses (ACH) to demonstrate a comprehensive set of design functionalities offered with our tool. 
Second, we present several case studies to indicate the usefulness of \name{} for automatically and semi-automatically generating compelling infographics.
Finally, we conducted a user study with 10 participants, including design novices and experts, to show the creative support of our framework. Further following the user study, we interviewed two expert designers in-depth to collect feedback on our framework and \name{}.


In summary, our main contributions in this paper include:
\begin{itemize}
    \item A general, extensible framework for infographic design that captures standard design components and workflows;
    \item An interactive tool, \name{}, that implements the framework to provide automatic and semi-automatic generation of infographics based on flexible user inputs and manipulation
    \footnote{We plan to open-source our tool, and the access will be provided upon the publication of this paper.};
    \item A dataset of 1k infographic images with labeled Visual Groups, and extracted individual Visual Group design SVGs and connection SVGs \footnote{The access to the dataset will be provided upon the publication of this paper. Several samples can be found in our supplementary materials.};
    \item Results and analyses from a three-part evaluation of our approach on various aspects.
    
\end{itemize}

\section{Related Work}
\label{s:related_work}
 
\subsection{Infographics and Relevant Studies}
\label{ss:infographics}
Infographics have recently acquired attention due to numerous studies highlighting their effectiveness in impacting its viewers compared to textual or even chart-based data presentation. The design of infographics depends on several factors related to human perception of information, which makes them more memorable and engaging \cite{bateman2010useful,haroz2015isotype,harrison2015infographic}. Some previous works have focused on understanding the use of visuals in infographics and their role in making the infographics more memorable and visually pleasing \cite{borkin2013makes,harrison2015infographic,skau2017readability}. For example Bateman \textit{et al.} \cite{bateman2010useful} discussed infographic design in terms of increased memorability. Haroz \textit{et al.} \cite{haroz2015isotype} studied infographics to assess the speed of finding information, user engagement, and information retention. Harrison \textit{et al.} \cite{harrison2015infographic} evaluated how engagement and memorability are determined in an infographic design. Similarly, Krum \textit{et al.} \cite{krum2013cool} mentioned the importance of initial engagement in the form of The Five-second Rule.

Studies have also been conducted to summarize an infographic design automatically. For example, Bylinskii \textit{et al.} \cite{bylinskii2017understanding} and Madan \textit{et al.} \cite{madan2018synthetically} studied infographic design by automatically extracting the contents of an infographic in the form of textual and visual elements using deep learning. Fosco \textit{et al.} \cite{Fosco2020} applied deep learning to predict the importance of different portions of a general graphic design.
Chen \textit{et al.} \cite{chen2019towards} analyzed timeline infographics and proposed some design component concepts specific to timelines, such as main body, event marks, and event annotations.
More generally, Lu \textit{et al.} \cite{lu2020exploring} summarized infographic designs based on the notions of Visual Information Flow (VIF) and Visual Group (VG). They clustered the infographic design space into 12 categories based on VIF; however, they did not fully investigate the design of VGs.

Although these studies help evaluate, understand, and categorize existing infographic designs, the goal of automated infographic generation is out of scope in all of these works, except Chen \textit{et al.}'s work that focuses only on timeline infographics. In this paper, we focus on automatically and semi-automatically generating infographics. Based on the above principles and concepts, especially the VIF and VG, we propose a flexible framework capturing the characteristics of infographics designs and generation workflows by decoupling the structure of general infographics into four key design components.   

\subsection{Infographic Generation Tools}
\label{ss:ig_tools}
Infographic generation tools can be classified into three main categories: manual, semi-automated, and fully automated tools. 
The manual techniques include design tools that provide complete control to designers for authoring every aspect of infographics from scratch, such as Adobe Illustrate and other commercial software \cite{photoshop, marvel, proto, mockflow, adobexd}. 

On the other hand, the semi-automated design tools allow for easier infographic generation while keeping some partial control to designers in the process. For example, tools like Proto.io \cite{piktochart}, and Timeline Storyteller \cite{brehmer2019timeline} support the generation of timeline infographics given custom time-series data from existing templates. 
Chen \textit{et al.} \cite{chen2019towards} improved the structure of timeline infographics by automatically extracting templates from infographic images using deep learning. 
Other works in this category include generating info-images, which overlays charts onto images to convey the statistics visually. For example, Infomages \cite{coelho2020infomages}, and Graphoto \cite{park2018graphoto} allow users to add and modify chart designs overlayed onto an image. Focusing on a different target, InfoNice \cite{wang2018infonice}, DataInk \cite{xia2018dataink}, and DataQuilt \cite{zhang2020dataquilta} support creating visually appealing pictorial charts or graphs from data by facilitating data mark authoring by users. 

Finally, the third category is fully automated tools that directly generate infographics from input data or resources, such as the ``Design Idea'' function of Microsoft PowerPoint. Also, DataShot automatically generates fact sheets based on existing templates by automatically extracting data facts from a custom dataset \cite{wang2019datashot}. Another work by Cui \textit{et al.} \cite{cui2019text} produces infographics automatically from text using pre-existing templates for simple proportion-related descriptions. 

Although all of these works provide different control levels to the designers, there are limitations to each category. The manual tools have a steep learning curve and require a high level of expertise to design infographics from scratch, hindering the adoption of novice designers. 
The semi-automated approaches focus on automating low-level designs, making it easier for novice designers to create infographics, but they only support either a particular class of infographics or specific pieces within an infographic. 
On the other hand, the fully automated techniques have the fastest turnaround time for generating infographics, but they are limited by prescribed whole-infographic templates, restricting users' freedom. 
We take a holistic view of infographic designs and propose an extensible framework, bridging the gap between semi-automated and fully automated techniques. 
Our \name{}, based on the framework, allows individual aspects of an infographic design, optimized and manipulated independently. Thus, designers can create unique infographics from simple markdown text input with both high-level (e.g., VIF layout) and low-level control (e.g., connecting elements), not limited to existing templates or categories.

\subsection{Visual Data Storytelling and Organization}
\label{ss:vio}
Visual storytelling plays a crucial role in presenting information in various formats. For example, Kosara \textit{et al.} \cite{kosara2013storytelling} put visual information organization, also known as storytelling, on the same level of significance as the exploration analysis step in visualization. Siegel and Heer \cite{segel2010narrative} classified visual information methods into seven formats such as animation and comics. Following up on this, for slideshow-style presentations, Hullman \textit{et al.} \cite{hullman2013deeper} analyzed 42 narrative visualizations to gain empirical knowledge to pick the best sequence of slides. Similarly, Bach \textit{et al.} \cite{bach2016telling} analyzed the visual information organization in comics, further introducing graph comics as a storytelling medium. Kindlmann \textit{et al.} \cite{kindlmann2014algebraic} introduced the three general principles for a good visualization design. Later these principles were incorporated in a mathematical model for the design of information visualization by Moere \textit{et al.} \cite{moere2011role}.

Many tools have recently been developed to support the visual organization of information for storytelling in different mediums. For example, Kim \textit{et al.} \cite{kim2016data} and Liu \textit{et al.} \cite{liu2018data} designed a framework to create graphical elements with data binding. 
ChartAccent \cite{Ren2017ChartAccent} supports users to create various annotations on data charts for storytelling manually.
Lee \textit{et al.} \cite{BongshinLee2013SketchStory} proposed SketchStory, designed for a large whiteboard that recognizes certain sketch gestures for chart creation.
Kim \textit{et al.} \cite{Kim2019DataToon} developed DataToon for authoring data comics for dynamic networks with pen+touch interactions.
TimeLineCurator \cite{fulda2015timelinecurator} features the authoring of timeline visualizations by extracting event information from unstructured text. 
DataClips \cite{amini2016authoring} allows novices to assemble data-driven clips together to form a longer video effectively.

While one usage of infographics is to support storytelling, most of the existing research in this area focuses on specific genres such as charts, comics, data videos, and timelines, which is out of the scope of generating common infographics. 
On a related aspect, the VIF represents the story exhibited in the data. Our \name{} supports storytelling to some extent by recommending VIF layouts from user input and allowing users to sketch a VIF on a canvas (with or without a custom pivot graphic). It also helps novices focus on the story content rather than the design by supporting a Markdown-based input. 


\subsection{Visualization Recommendation}
\label{ss:vis_reco}
Besides the work related to infographic generation, there is considerable research in developing recommendation systems for data visualizations (charts). These techniques can be broadly classified into two categories, rule-based techniques and data-driven techniques. 

The rule-based techniques include initial work by Steele \textit{et al.} \cite{steele2011designing} who narrowed down rules for developing visual encodings for visualizations, which could then be used for ranking. However, these theoretical rules are not always practical as there is dependence on the users' various constraints, as mentioned by Moritz \textit{et al.} \cite{moritz2018formalizing} in DRACO. They propose visualization to be done based on constraints along with theoretical concepts.  These concepts were further extended with APT \cite{mackinlay1986automating} which introduced methods to generate a latent space of charts using compositional algebra, which was later improved by SAGE \cite{roth1994interactive}. To support efficient search in this latent space of charts, CompassQL \cite{wongsuphasawat2016towards}, later improved by Voyager \cite{wongsuphasawat2015voyager} and Voyager 2 \cite{wongsuphasawat2017voyager} were developed using specific query specifications on this search space. 

In the data-driven techniques, VizML \cite{hu2019vizml} uses machine learning to develop encodings representing the relationship between data characteristics and visualizations. These encodings can then be used to recommend visualizations for a given dataset. Data2Vis \cite{dibia2019data2vis} is another deep learning framework to directly generate visualizations given the data using the visualization grammar introduced in Vega \cite{satyanarayan2015reactive} and VegaLite \cite{satyanarayan2016vega}. There also exist techniques that combine rule-based methods and data-driven machine learning methods, such as DeepEye \cite{luo2018deepeye} to rank and classify visualizations. 

While the above systems are suitable for ranking data visualizations, they only work for charts and do not support infographics ranking. Infographics have significantly different characteristics than standard charts, and generating and ranking them is inherently different from working with charts, which is the focus of our work.

\section{Infographics Generation Framework}
\label{s:framework}

\begin{figure}[tb]
 \centering
 \includegraphics[width=\columnwidth]{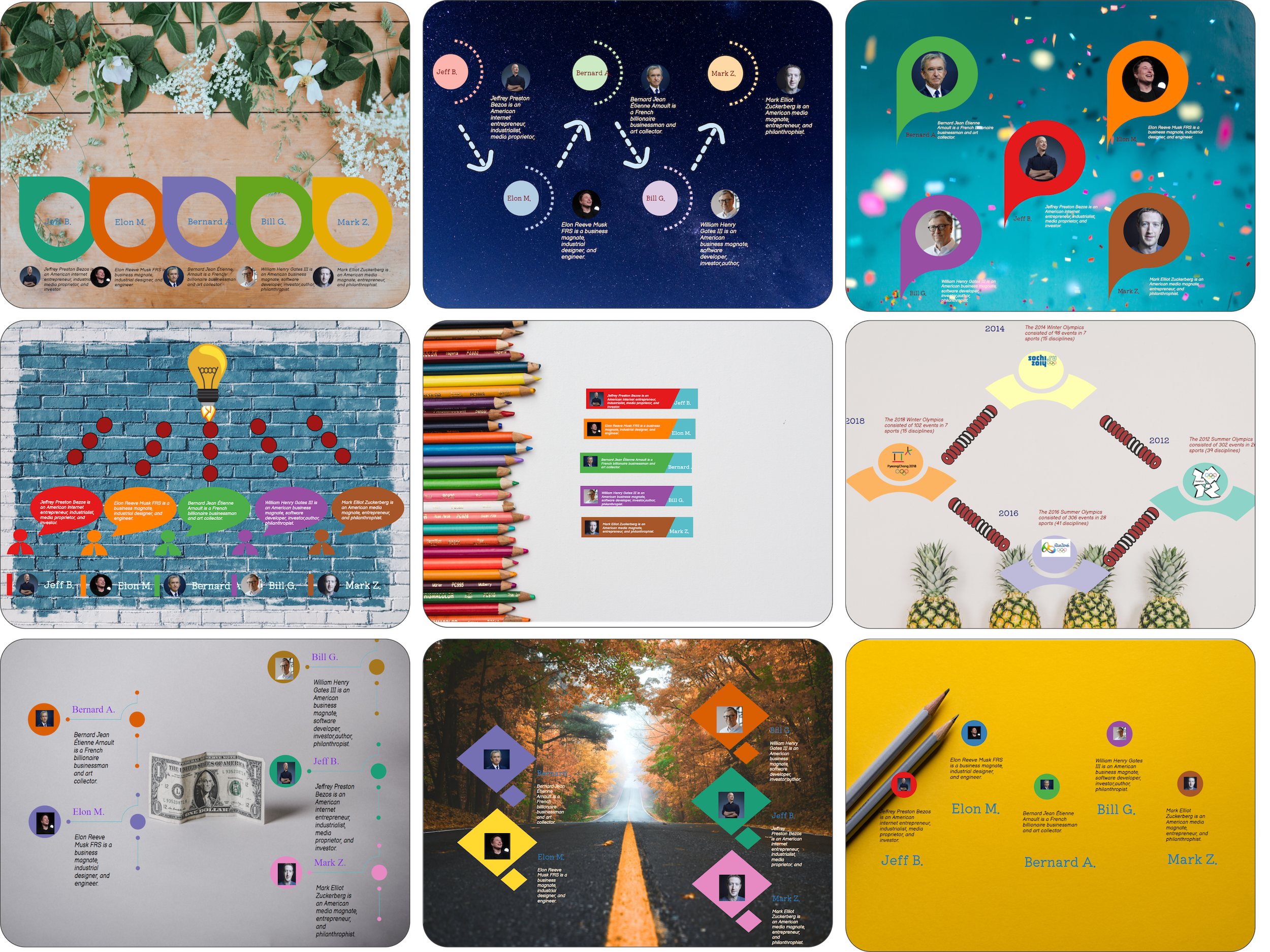}
 \caption{Sample infographic designs generated with our framework. These infographics were collected during a user study discussed in Section \ref{s:evaluation}.}
 \label{fig:collage}
\end{figure}

In this section, we introduce our flexible framework for infographics generation, which is distilled from a formative study with participants with different levels of experience in infographic design. Example infographics generated using our framework are shown in Figure \ref{fig:collage}.

\subsection{Formative Study}
\label{ss:survey}

To systematically evolve our idea of an infographic generation framework, we first conducted a formative study to get to know designers' requirements, their views of infographic designs, and general workflows. 
This approach helped concertize our framework and tool design with a user-centered evaluation at an earlier development stage. 

The formative study participants were carefully chosen to be designers and researchers working in data visualization and infographic design, with various experience levels. Out of ten participants, two were professional designers working in the industry, two were professors working in data visualization, three were Ph.D. students working in data visualization, and three were undergraduate students in Computer Science interested in information visualization.

The participants were initially introduced to infographics and the target we try to achieve. We asked how they would conceptually model infographic designs. Next, they were introduced to the VIF and VG concepts, followed by discussions about existing infographic generation tools and their shortcomings. 
We further inquired how they would use the existing tools to design infographics and the ideal tools they could imagine. 
This helped us devise a new framework that could flexibly support automated infographic generation and recommendations for various cases. 

\subsection{Key Infographics Design Components}
\label{ss:components}

\begin{figure}[tb]
 \centering
 \includegraphics[width=\columnwidth]{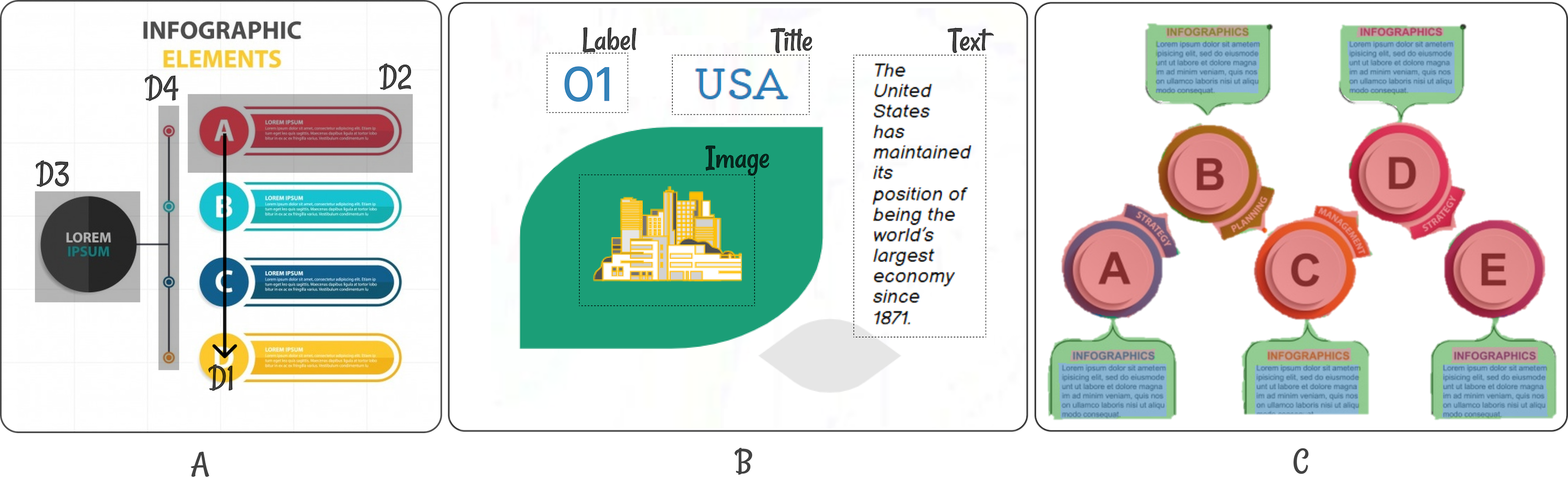}
 \caption{\textbf{(A)} Design components of an infographic: \textit{D1} is the Visual Information Flow (VIF) layout, \textit{D2} is the Visual Group (VG) design, \textit{D3} is the pivot graphic, and \textit{D4} are the connecting elements. \textbf{(B)} Individual components inside a VG. \textbf{(C)} Example of segmented VGs in our infographics dataset. Each VG and their individual graphic elements are segmented by human annotators.}
 \label{fig:study_vg_dataset}
\end{figure}

Based on the study, we reconfirmed the importance of the VIF and VG concepts proposed in Lu \textit{et al.}' work \cite{lu2020exploring}.
We further consolidated four design components through designing a new framework, shown in Figure \ref{fig:study_vg_dataset}A. Each of the components is critical for designers to control for generating customized infographics. Various levels of controls result in a flexible workflow with fully- or semi-automated tools. The four key design components are:

\textbf{D1: Visual Information Flow Layouts.} 
The participants confirmed that the concept of VIF reflects the backbones of the story or information that an infographic aims to express. Lu \textit{et al.} \cite{lu2020exploring} discovered 12 common VIF layouts. Participants further agreed that allowing designers to optionally control the VIF layout intuitively is an essential component of any infographic generation workflows.

\textbf{D2: Visual Groups Designs.} 
The participants also echoed the concept of VG, and they stated that VGs could be designed separately and reused in one or multiple infographics. These VGs could then be placed according to the VIF to generate an infographic. 
They also mentioned that it is essential to mitigate the effort in VG design for novices while allowing them to fully control the content of VGs and the overall structure of an infographic. The participants further mentioned that an interface separating the content and the design, like CSS and HTML for web design, could be of great benefits.  

\textbf{D3: Pivot Graphics.} 
The participants pointed out that some infographics contain a pivot graphic, which are design elements different from VGs, acting as a central background component or binding anchor for an infographic. This coincided with the ``main body'' of timeline infographics in Chen \textit{et al.}'s analysis \cite{chen2019towards}. Participants observed that such pivot graphics exist in a wide range of infographics in addition to timelines and stated that supporting the customization of this element is needed in infographics design when appropriate.

\textbf{D4: Connecting Elements.} 
The participants formulated the connections between VGs, sometimes between a pivot graphic and different VGs, as one of the critical infographics components. The connecting elements cannot be included in VGs because they are decorations not contributing to the content semantics of the VGs, and sometimes there is no one-to-one mapping between the connections and the VGs. 
To support infographic generation, we categorized the connecting elements into four classes: flow-shaped, regular, alternating, and pivot connections, based on our discussion with the participants (see Section \ref{s:connection} for details).

\subsection{Infographics Generation Pipeline}
\label{s:overview}

\begin{figure*}[tb]
    \centering
        \includegraphics[width=\columnwidth]{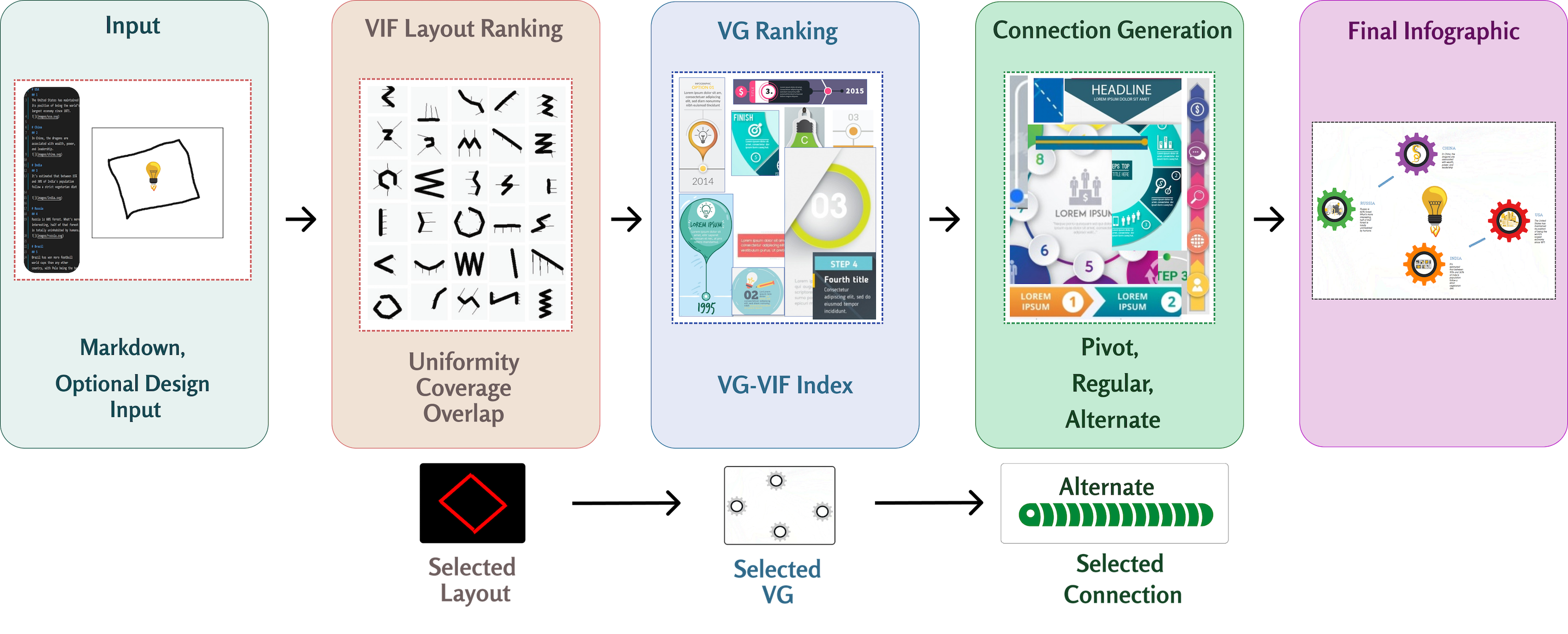}
    \caption{The three-stage pipeline of our framework, discussed in detail in Section \ref{s:infographic_generator}. The first stage recommends the VIF layouts using an energy function from given input information and optional layout or pivot elements. In case the user provided a hand-drawn layout as input, we match the existing VIF layouts to this design and rank them based on similarity to the sketch. In the second stage, we select the VG designs using our \textit{VG-VIF} index. Finally, in the third stage, connections and colors are generated to provide a final touch to the infographic design.}
    \label{fig:pipeline}
\end{figure*}

Based on the design components discussed above, we construct a pipeline within the framework with three main stages, as shown in Figure~\ref{fig:pipeline}. Input to the first stage is the user-provided information, which is used to decide the number of VGs required to create an infographic. Shown as \textit{Input} in Figure~\ref{fig:pipeline}, the information can be provided in the form of markdown text, where each bullet point can include a title, text, label, and an image. The user can optionally provide the pivot graphics like images and layout drawings using a canvas-based interface. This information is then passed onto the following stages:

\textbf{S1: Fitting Visual Information Flow Layouts.} 
This is the first stage where the existing VIF layouts extracted from infographic datasets are ranked based on the pivot graphics (if any), canvas size, and the user's sketch input (if any) to recommend the best fitting positions to place the VGs. We discuss the details of our curated dataset in Section \ref{s:dataset} and the ranking algorithm in Section \ref{s:infographic_generator}. 

\textbf{S2: Selecting Visual Group Designs.} 
Once the VIF layouts are ranked, the next step is to select the VG designs from existing datasets, which best fit the proposed VIF layouts. We curated a VG-VIF index for this purpose, explained in Section \ref{s:infographic_generator}, providing final VG designs that can generate the infographic. 
The user can alter the proposed designs by choosing from multiple recommendations or going back to the previous stage to provide new inputs. Once the VG design is finalized, it is placed based on the VIF layout, rotated, and scaled to fit the infographic's size and design. The details of rotation and scaling of VGs are discussed in Section \ref{s:infographic_generator}.

\textbf{S3: Generating Connections and Styling.} After finalizing the VIF layouts and VG designs, the final step is to connect the pivot elements (if any) and the VGs and stylize the VGs using a pattern to give the infographic a final touch. The algorithm to generate connections and colors is explained in Section~\ref{s:infographic_generator}. 

Using this three-step process, the final infographic design recommendations are generated automatically with optional VIF, pivot element, or VG design input, thus bridging the gap between the existing fully- and semi-automated infographic generation techniques. 
Novice designers can take advantage of the fully automated design pipeline to explore a vast set of infographic design recommendations, while more experienced designers can use our framework to explore a set of designs possible based on some design constraints. Designers can also export the generated infographics as SVG files to fine-tune and alter very low-level design elements.  
\section{Infographics Dataset with Visual Groups}
\label{s:dataset}


As mentioned previously, our framework, or any automated or semi-automated infographics generation tools, need to be driven by existing infographic designs created by experts, thus adequately leveraging the collective wisdom, aesthetics principles, and perceptual rules derived by professionals.
However, there still a lack of high-quality infographics datasets with fine-grained and accurate annotations. 
The infographics dataset of 1K images with annotated VG designs is one of our main contributions in this work. The main goal to curate this dataset was to compensate for the lack of hand-designed VG datasets. Since VGs in infographics are a reasonably new idea, recently introduced by Lu \textit{et al.} \cite{lu2020exploring}, there does not exist any dataset targeting specifically towards annotating the VGs in infographics. Creating this dataset helped us extract VG designs to extend the scope of infographic designs that can be generated with our framework. 

Along with the 1K human segmented VGs in complete infographic images, we also release the extracted VGs from each of these images in separate SVG files. These designs are used in our framework to generate infographics. We annotate the VGs with both a segmentation mask and a bounding box, while we only segment the individual VGs since the bounding box information of these components is already available in the existing datasets \cite{lu2020exploring}. The process of data collection is discussed in the following text, and a sample of the collected dataset is provided with the supplementary material.

\subsection{Experiment Setup}
We annotated the infographics using Amazon Mechanical Turk \cite{mturk}. The workers were first introduced to the concept of VGs and their individual components with some example images, as shown in Figure~\ref{fig:study_vg_dataset}B. The task was to segment individual VGs, the components inside the VGs, and annotate VGs with a bounding box in each of the infographic images. For consistency, each image was annotated by three different workers, and we chose the best annotated VGs for each image manually. An example user annotated image is shown in Figure~\ref{fig:study_vg_dataset}C.


\subsection{Source Images}
We used the study results by Lu \textit{et al.} \cite{lu2020exploring} to sample 1000 infographics belonging to each of the 12 VIF categories and in the ratio as they appear in the dataset. The number of images in each VIF category used in the dataset is provided in the supplementary material. Because of our framework's design aspect, since we aimed to relate the VG designs with VIF layouts, sampling infographic images based on the VIF layout was crucial for consistency with real-world designs.  

\subsection{Processing the Segmentation Maps}
Human-generated segmentation maps are coarse with ill-defined shapes. To generate a well-defined VG segmentation map, we employed the automated GrabCut \cite{rother2004grabcut} algorithm, similar to \cite{chen2019towards}. The human annotations were used as input to the algorithm to generate high-quality segmentation masks for VGs. This mask was then passed as input to Solaris \cite{solaris} which generated SVG paths for segmented VG designs. Finally, we added additional components to the SVG based on bounding box annotations of individual VGs, provided with the dataset by Lu \textit{et al.} \cite{lu2020exploring}.

\section{Design Components Characterization} 
\label{s:design_study}

Our framework's efficiency highly depends on the quality and quantity of existing VIF and VG designs. Since we generate infographic designs by ranking the existing VIF and VG designs, we created a dataset of VIF layouts and VG designs by expanding upon the design components discussed in Section \ref{ss:components}. In this section, we discuss the process of extracting these components from the existing infographics dataset. 

\subsection{Visual Information Flow Extraction}
VIF is the first component of our framework (\textbf{D1}), representing the backbone layout inside an infographic. As shown in Figure \ref{fig:study_vg_dataset}A, VIF represents the direction of information flow following the positions of repeated VGs inside an infographic. It is represented as a series of 2D points where each point represents a VG position in two dimensions corresponding to a design area, and the number of points represents the number of VGs in that particular flow. The sequence of points represents the sequence of visual groups based on the information flow. To calculate the VIF for a given infographic, we used the algorithm discussed in \cite{lu2020exploring}. 

\subsection{Visual Group Design Extraction}
Since VG designs are the basic building block for infographic generation in our framework (\textbf{D2}), we chose multiple sources to collect VG designs to attain maximum coverage. Out of the total three sources we used, the first one is Adobe Stock \cite{adobe}, which contains SVG designs of several human-generated infographics, from which we manually separated the VG designs; the second source is our human segmented VG dataset which we curated for this work, explained in Section \ref{s:dataset}; the third source is taken from the work by Chen \textit{et al.} \cite{chen2019towards} where we extracted VGs from timeline infographics using a Mask R-CNN \cite{he2017mask}, explained in Section \ref{ss:vg_maskrcnn}. 
These sources were carefully chosen so that we get the VG designs created directly by humans and also using automated deep learning networks. We chose the SVG format to store the VG designs because of various benefits like scaling, color, and widespread usability. Each VG is further divided into four components as shown in Figure \ref{fig:study_vg_dataset}B, namely: \textit{Label}, \textit{Text}, \textit{Title}, and \textit{Image}. Each VG can contain any number of these elements, and each SVG design contains separate markings for each component, which the user's content can then replace as per our design pipeline.

\subsubsection{Mask R-CNN for Initial Element Segmentation}
\label{ss:vg_maskrcnn}

\begin{figure}[tb]
    \centering
        \includegraphics[width=\columnwidth]{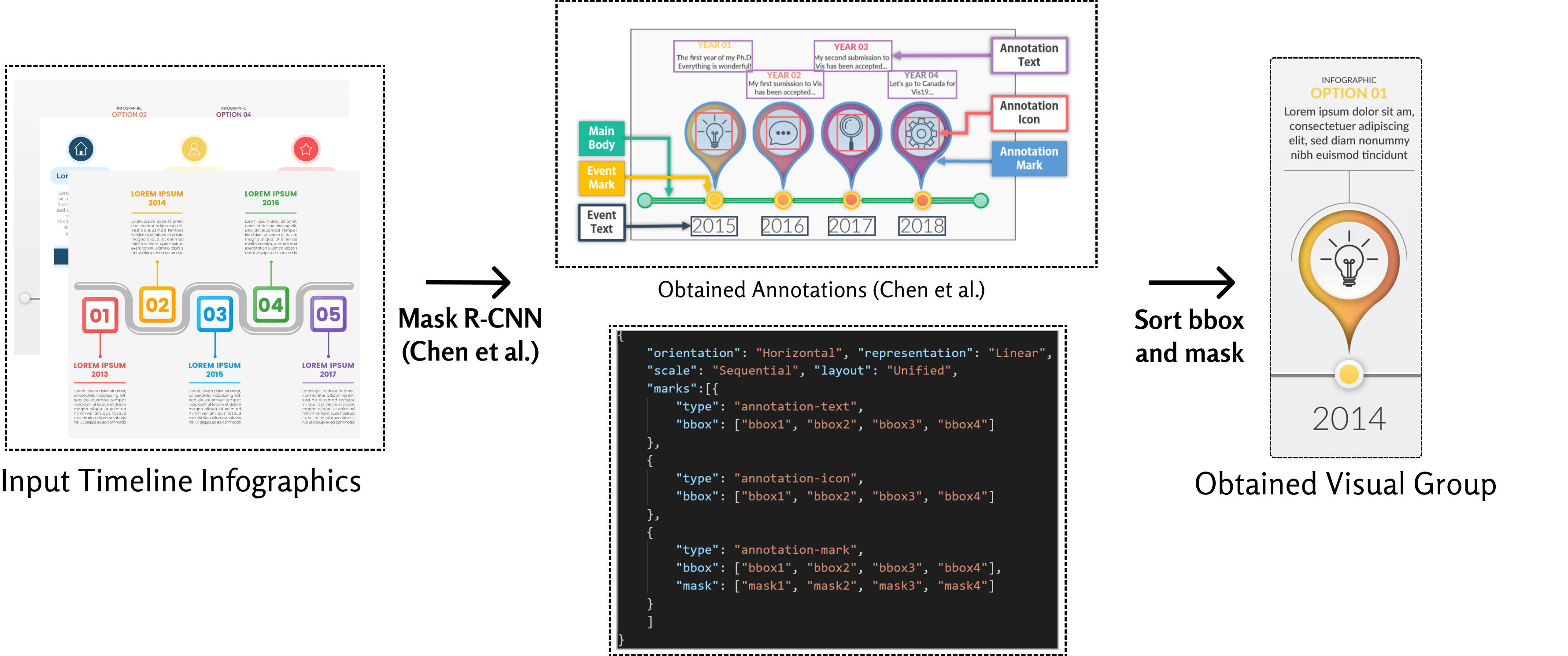}
    \caption{Process of generating VG designs from the Mask R-CNN network. After generating the network's annotations, the bounding box and mask values are sorted in the direction perpendicular to the predicted orientation. Then the first values of every bounding box and mask are grouped to get a VG.}
    \label{fig:mask_rcnn}
\end{figure}

In addition to the VG designs generated from the datasets available at Adobe Stock \cite{adobe}, we modified the timeline infographics work by Chen \textit{et al.} \cite{chen2019towards} to extract more VG designs. The timeline infographics dataset released with this work contains 9592 manually created timeline infographics from Timeline Storyteller \cite{brehmer2019timeline} and 1138 infographics from online sources. It includes 4689 images with manually labeled infographics components. 

We used Mask R-CNN \cite{he2017mask} with this dataset to generate the segmentation maps for the infographic components. Besides predicting the segmentation masks, a modified network predicts the local and global information about a timeline infographic image. 
The global information predicted by the network is the infographic \textit{Representation}, \textit{Scale}, \textit{Layout}, and \textit{Orientation}. For example, an infographic can have the global properties \textit{Linear}, \textit{Sequential}, \textit{Unified}, and \textit{Horizontal}, as shown in Figure \ref{fig:mask_rcnn}. Besides the global information, the network also predicts the bounding boxes for local information like \textit{Text}, \textit{Label}, \textit{Title} and \textit{Icon} along with the segmentation masks for the designs containing these elements. Figure \ref{fig:mask_rcnn} shows the network and sample annotations, respectively. The complete details of these global and local properties can be found in \cite{chen2019towards}. 

As we can see, the output of the above model provides us with the annotation labels of \textit{Text}, \textit{Icon} and \textit{Mark} and event labels of \textit{Mark} and \textit{Text} along with \textit{Main Body} labels. We can convert each of these labels to the corresponding VG elements based on the relations shown in Table \ref{tab:relations}. The entire pipeline is shown in Figure \ref{fig:mask_rcnn}.

\begin{table}[tb]
\caption{Relations of VG Elements with the Timeline Infographics prediction from Mask-RCNN}
\label{tab:relations}
\begin{tabu}{
    r
    *{2}{c}
    *{5}{r}
}
\toprule
  \textbf{Timeline Component} & \textbf{Visual Group Component} \\
  \midrule
  Annotation Text & Text and Title \\
  Annotation Icon & Icon \\
  Annotation Mark & Visual Group Design \\
  Event Mark & Connecting Point \\
  Event Text & Label \\
 \bottomrule
\end{tabu}
\end{table}

\subsubsection{Visual Group SVG Generation}
Creating reusable VG designs from the Mask R-CNN predictions requires binding several predicted bounding boxes based on their respective VGs. Since the Mask R-CNN output does not predict which VG a particular bounding box belongs to, we developed a simple scheme to bind the bounding boxes based on their respective VG positions. Also, since the purpose here is to get the VG design, we only created one VG component from each infographic. We assume that the design of the VGs remains consistent within the infographic, which was found to be true in all the timeline infographics from the dataset by Lu \textit{et al.}

Hence, after predicting various elements for a given infographic image, we used the \textit{Orientation} as predicted by the Mask R-CNN to generate the VG design. We assumed that VGs are spread in the direction perpendicular to the infographic orientation. For example, in Figure \ref{fig:mask_rcnn}, the VGs extend in a vertical direction when the infographic is horizontal. Using this assumption, after generating the marks from Mask R-CNN for each of the infographic design elements, we sorted every element based on the bounding box centers in the direction perpendicular to the predicted orientation. 
After, we assigned the first element of each predict mark to the same VG. 

For simplicity, to solve the purpose of extracting meaningful VG designs, we only used the infographics with \textit{Horizontal} and \textit{Vertical} orientations. Also, we confirmed the extracted VG designs by a human analyst for their correctness and discarded the incorrectly generated VGs. The final bounding boxes and masks obtained from this process were converted to the SVG-based VG designs using Solaris \cite{solaris}.

\subsection{Connecting Elements Characterization}
\label{s:connection}

\begin{figure}[tb]
    \centering
        \includegraphics[width=\columnwidth]{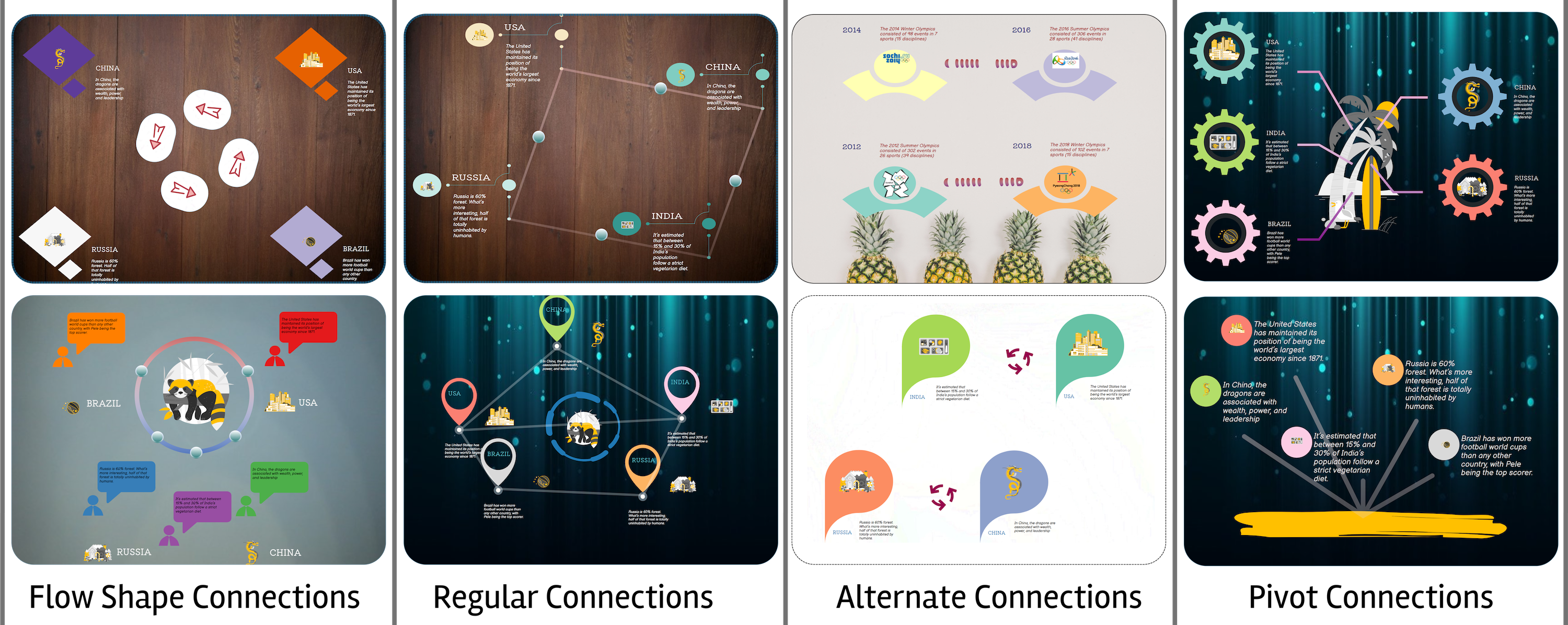}
    \caption{Four types of connection styles that can be generated with our framework.}
    \label{fig:connections}
\end{figure}

The connections in an infographic refer to the design components used to connect VGs and pivot graphics with respect to the VIF layout (\textbf{D3, D4}). Based on our formative study, we collected SVG connection designs from various online sources. Since the ``connection styles'' are mainly determined by the layouts and designs of components such as the VIF and VGs, we categorize them for a given infographic into the following four classes, as shown in Figure \ref{fig:connections}:

\textbf{Flow Shape Connections:} These connections are placed around the center of the infographic or around the pivot element (if present). Following the direction of the infographic's VIF layout, their placement is decided based on the placement of VGs on the VIF flow, and the slop is decided by the slope of the corresponding flow line, connecting two VGs.

\textbf{Regular Connections:} These connections follow the direction of the VIF layout of the infographic and are placed at the center of each flow line. The placement angle follows the slope of the flow line and the length of each connection depends on the distance between corresponding VGs in the infographic. 

\textbf{Alternate Connections:} These connections are similar to regular connections but are placed at alternating flow lines. Like regular connections, the angle and position of these connections depend on the flow line's center and the slope, and the length depends on the distance between VGs. 

\textbf{Pivot Connections:} In case the infographic has some pivot graphics, these connections are generated from the center of the pivot element towards all the VGs. For a particular VG, the connection placement is in the center of the line joining the pivot element and the VG. Also, the connection angle is the slope of connecting line, and the length is decided based on the distance of the VG from the pivot element. 

\section{Infographics Wizard}
\label{s:infographic_generator}

Following the indicated pipeline of our framework (see Section \ref{s:overview}), we developed an interactive tool, \name{}, that provides infographic recommendations of different design components in parallel for rapid prototyping and exploration of infographics. An overview of the tool is shown in Figure \ref{fig:teaser}. 
This section discusses the algorithms for selecting the best fitting VIF layouts and VG designs in more detail. We also include the details of information organization inside VG designs, VG rotation and scaling based on a VIF layout, and connecting elements generation. 
Note that \name{} is a specific realization of our framework, where other suitable algorithms and datasets can be employed to fulfill the proposed pipeline. 

\subsection{Visual Information Flow Layout Recommendation}

\begin{figure}[tb]
    \centering
    \includegraphics[width=\columnwidth]{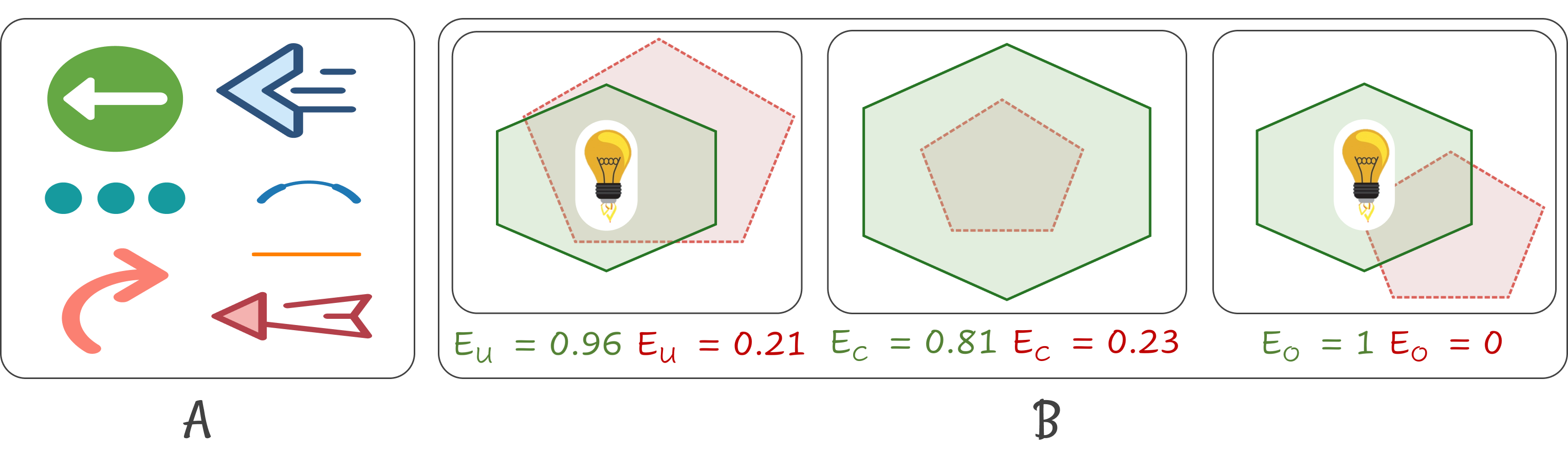}
    \caption{(A) Example connection designs in our dataset used to generate connections. (B) The VIF layouts are ranked based on three energy functionals. $E_{U}$ gives the score to uniformity, meaning how uniformly the VIF vertices are located from the pivot element center. $E_{C}$ scores the coverage based on how much the convex hull of a VIF covers the canvas area. $E_{O}$ is the overlap score, which is $0$ if there is an overlap between any VIF vertex and the pivot element, and $1$ otherwise.}
    \label{fig:layout}
\end{figure}

This algorithm finds the best layouts from our extracted VIF layouts dataset, given the user design constraints. We did not optimize for layout generation directly and instead chose a ranking-based approach because retrieval of related existing VIF layouts is faster and less resource-intensive than directly optimizing for finding the best layout \cite{o2015designscape}. Using the ranking-based approach, we can recommend relevant layouts in real-time and instantly following user feedback. The ranking of VIF layouts are scored based on the energy functional shown in Equation~\eqref{eq:layout_rank} by $E_{L}$
\begin{equation}
    \label{eq:layout_rank}
    E_{L} = E_{O}(\alpha E_{C} + (1-\alpha) E_{U})
\end{equation}
where $\alpha \in [0,1]$, balances the contribution of $E_{U}$ and $E_{C}$ in the final energy functional. $E_{U}$, $E_{O}$ and $E_{C}$ are the three components of the energy functional, with the intuition shown in Figure \ref{fig:layout} (B). $E_{O}$ is an indicator function that is $1$ if and only if none of the VIF flow elements lie in the boundaries of the bounding box of the pivot element, and $0$ otherwise. 

Another component of the layout scoring energy functional is $E_{C}$, which calculates the score based on the coverage a VIF layout provides. More the coverage, the higher the value of $E_{C}$, as shown in Equation~\eqref{eq:em}. Here $Hull(VIF)$ means the convex hull of the points in a particular VIF layout. 
\begin{equation}
    \label{eq:em}
    E_{C} = \frac {Area\ of\ Hull(VIF)}{Area\ of\ Canvas}
\end{equation}

The last component of the layout scoring functional is the uniformity in the distance of the VIF elements wrt to the center of the pivot elements. To calculate this, we first calculate the center of the bounding box of the pivot element, denoted as $P_{XC}$ and $P_{YC}$, shown in Equation~\eqref{eq:pxc} and \eqref{eq:pyc}.
\begin{equation}
    \label{eq:pxc}
    P_{XC} = P_{X} + \frac{P_{L}}{2}
\end{equation}

\begin{equation}
    \label{eq:pyc}
    P_{YC} = P_{Y} + \frac{P_{W}}{2}
\end{equation}

Then, we calculate the variance of the distance between each VIF element and the pivot center, which gives a score of uniformity in the spread of the layout, shown in Equation~\eqref{eq:eu}. Lesser the value of $E_{U}$, the better the uniformity and vice versa. $\bar{D}$ is the mean Euclidean distance between the VIF elements and the pivot center. 
\begin{equation}
    \label{eq:eu}
    E_{U} = \frac {\sum_{X_{i}, Y_{i} \in VIF} \sqrt{(X_{i}-P_{XC})^2 + (Y_{i}-P_{YC})^2} - \bar{D}}{N}
\end{equation}

In some cases, designers might provide an initial sketch of a coarse layout, for example, shown in Figure \ref{fig:teaser} and \ref{fig:pipeline}. In such cases, since we already have a VIF layout, we aim at finding the nearest neighbors to this user-provided VIF layout from our dataset. To calculate the nearest neighbors, we first calculate the extreme points on the hand-drawn contour using the technique by Teh \textit{et al.} \cite{teh1989detection}. These points give us the estimated positions of VGs, which are then matched with the existing database of VIF layouts to find the closes neighbors of these extreme points. These neighbors give the ranking of VIF layouts wrt to a hand-drawn contour provided by the designers.

\subsection{Visual Group Design Recommendation}
\label{ss:vg_reco}
Based on our infographic generation pipeline, after ranking the VIF layouts, the next step is to choose corresponding VG designs that best fit the selected VIF layout. To support this, we develop a \textit{VG-VIF index} that provides a method to rank VGs given a VIF layout from our dataset.

\subsubsection{VG-VIF Index}
\label{ss:vg_vif}
The VG-VIF index aims to capture a global relationship between the VG designs and the VIF layouts to support a ranking system that allows accurate recall of the suitable VG designs given a VIF layout. To develop the VG-VIF index, we first clustered all the VIF flows in our dataset into 12 categories as mentioned in \cite{lu2020exploring}. We first generated the VIF images for each infographic in our dataset to generate these clusters. These images were then reduced to 50 dimensions using PCA \cite{joliffe1992principal} to get the principal components, which were then further reduced to 2D using t-SNE \cite{van2008visualizing}. This 2D space was divided into clusters with centers around the high-density regions with DBSCAN \cite{liu2006fast} and then iteratively clustering all the points into one of the centers. We repeated this procedure until there was no further improvement in the forming clusters. We used the same 12 classes discussed in \cite{lu2020exploring} to form the basis of the VIF layouts (denoted as CL-$X$ in Table \ref{tab:vg-vif}). 

Following this, we extracted the VG designs from each of the infographics. Table \ref{tab:vg-vif} shows an example of $4$ Visual Groups (denoted as VG-$x$) that can occur in multiple VIF flow categories (denoted as VIF-$x$). This example shows the one-to-many relationship between the VG designs and the VIF layouts. We can treat these relations as ``words'' occurring in some ``documents'', where the VGs are the documents and the corresponding cluster centers are the words in each document. 
Using this analogy, we obtain the ranking of VGs given a VIF cluster ID by calculating the \textit{TF-IDF} score \cite{ramos2003using} for each VG. The TF-IDF score has been extensively used in information retrieval and has properties that we can use to rank the VGs. Firstly, the TF-IDF score down-weights the most commonly appearing words; in this case, it results in higher rankings of less frequent VGs. Thus, our recommendation engine generates infographics using the VGs, which are less common. Also, the TF-IDF scores are generated covering the entire domain, in this case, the domain of VIFs, which is a desirable property while generating infographic designs.  

\begin{table}[tb]
\caption{A VG-VIF index stores the information of VGs, and the corresponding VIF cluster centers \textit{CL-$X$} in which they appear. Each VIF layout is clustered into one of the total twelve classes, shown as CL-$X$.}
\label{tab:vg-vif}

\begin{tabu}{
    r
    *{2}{c}
    *{5}{r}
}
\toprule
  \textbf{Visual Group ID} & \textbf{VIF Layout Cluster IDs} \\
  \midrule
  VG-1 & CL-1, CL-3, CL-4, CL-7 \\
  VG-2 & CL-1, CL-2\\
  VG-3 & CL-4, CL-5, CL-6\\
  VG-4 & CL-1, CL-2, CL-3, CL-4, CL-5, CL-6\\
 \bottomrule
\end{tabu}
\end{table}

To summarize the VG ranking process, we choose the best VIF layout vectors to place the VGs based on the canvas elements. The VG scores are obtained from the VG-VIF index, which shows how well-fitting a VG is for a given layout. The VG-VIF index scores are sorted, and a subset of high-scoring VGs are selected, which match the user's markdown input (i.e., the number and type of components inside a VG).

\subsubsection{Visual Group Placement and Information Embedding}
The ranking algorithm discussed above provides a list of VGs, which can be used with a selected VIF layout. However, placing the VG on the VIF layout requires proper rotation and scaling. The rotation of a VG is defined based on whether or not the infographic has a pivot graphic. In case the pivot graphic is present, each VG should face towards the pivot graphic, and hence the rotation angle is calculated by measuring the angle of the arc connecting the VG from the center of the pivot graphic on the circle centered at the pivot graphic. After the rotation angle for each VG is fixed, the input information can be embedded inside each VG. 

To embed the user input, each VG has placeholders defined in their respective SVG files for placing individual components (i.e., image, text, title, and label). Text's font size and dimensions of the image to be embedded are calculated based on the size of these placeholders and the content's size. Also, embedded text and images are invariant to the VG rotation and are always placed at the same angle in each of the VGs. 

\subsection{Connection Recommendation and Filtering}
Our connection recommendations are similar to the VG ranking algorithm, discussed in Section \ref{ss:vg_reco}. We categorize the connection ``styles'' into five classes, of which four were discussed in Section \ref{s:connection}, and 1 class ``none'' was added, which means that the infographic has no connecting elements. Like VGs, the goal is to rank these five classes of connection styles based on the VIF layout. For this purpose, we manually created a ``C-VIF'' index to store the connection styles and corresponding VIF layouts from chosen 200 infographics. These 200 infographics were carefully chosen to equally represent all the 12 VIF categories from Lu \textit{et al.} \cite{lu2020exploring}. Following this, we calculated the scores using the TF-IDF \cite{ramos2003using} scheme to generate rankings of connection styles given a VIF layout, as discussed in Section \ref{ss:vg_vif}.

Along with ranking connection ``styles'', we also collected SVG ``designs'' of connecting elements from various online sources, as shown in Figure \ref{fig:layout} (A). For each of these connection designs, we randomly picked the connection designs to be shown to the user on \name{}, shown in Figure \ref{fig:teaser} (D). The user can choose to browse more connection designs or pick a connection design, while the tool automatically controls the best connection styles to use in the recommendations. Although designers have the freedom to control the connection style still if required, using the ``Connection-Type'' option on the menu bar of \name{}, as shown in Figure \ref{fig:teaser}.


\section{Evaluation}
\label{s:evaluation}

We evaluated our framework and the interface \name{}, by following the schemes discussed in the nested model for visualization interface design \cite{munzner2009nested}. First, we use the Analysis of Competing Hypotheses (ACH) \cite{heuer1999analysis} to effectively compare the existing infographic generation tools with \name{} for design functionalities. Second, we present three case studies on \name{} to demonstrate that our framework generates high-quality infographic designs across various scenarios. Finally, to evaluate our framework and its interface for usability across designers with various expertise levels, we conducted a user study where we asked the participants to author and explore infographic designs using \name{} and rate the tool based on their experience across different usability aspects. We further interviewed two expert designers to collect detailed feedback. 

\subsection{Design Features Evaluation using ACH}
\label{ss:ach}

\begin{table}[tb]

  \caption{ACH, comparing \name{} and existing infographic generation tools based on the hypotheses formulated during the formative study, discussed in Section \ref{ss:survey}. A checkmark means the proposed hypotheses is satisfied by the tool and the cross mark means the hypotheses is not satisfied. The result row at the bottom shows the tools which satisfy all the proposed hypotheses.
  \label{tab:comp_hyp}}
  \begin{tabu}{%
	r%
	*{9}{c}%
	*{6}{r}%
	}
  \toprule
  \rotatebox{90}{Hypotheses} & \rotatebox{90}{Chen et al. \cite{chen2019towards}} & \rotatebox{90}{Timeline Storyteller \cite{brehmer2019timeline}} & \rotatebox{90}{Text-to-Viz \cite{cui2019text}} & \rotatebox{90}{VIF Flows \cite{lu2020exploring}} & \rotatebox{90}{DataShot \cite{wang2019datashot}} & \rotatebox{90}{Adobe Illustrator} \rotatebox{90}{MS PowerPoint} & \rotatebox{90}{Design Tools}  \rotatebox{90}{\cite{adobexd, photoshop, marvel, mockflow, piktochart, proto}} & \rotatebox{90}{\textbf{Infographics Wizard}}\\
  \midrule
  H1 & \checkmark & $\Cross$ & \checkmark & $\Cross$ & \checkmark & \checkmark & $\Cross$ & \checkmark \\
  \midrule
  H2 & \checkmark & \checkmark & $\Cross$ & \checkmark & $\Cross$ & $\Cross$ & \checkmark & \checkmark \\
  \midrule
  H3 & $\Cross$ & $\Cross$ & $\Cross$ & $\Cross$ & $\Cross$ & $\Cross$ & \checkmark & \checkmark \\
  \midrule
  H4 & $\Cross$ & \checkmark & $\Cross$ & $\Cross$ & $\Cross$ & $\Cross$ & \checkmark & \checkmark \\
  \midrule
  H5 & $\Cross$ & \checkmark & \checkmark & \checkmark & \checkmark & \checkmark & $\Cross$ & \checkmark \\
  \midrule
  \textbf{Result} & $\Cross$ & $\Cross$ & $\Cross$ & $\Cross$ & $\Cross$ & $\Cross$ & $\Cross$ & \checkmark \\
 \bottomrule
  \end{tabu}%
\end{table}

Analysis of Competing Hypotheses (ACH) \cite{heuer1999analysis} is the technique to evaluate a set of tools by comparing them through a set of hypotheses. Based on our application scenario, we compare the existing infographic generation tools listed in Section \ref{s:related_work} and \name{} in terms of design functionalities using ACH. The participants from the formative study proposed the hypotheses, discussed in Section \ref{s:design_study}, as they are a sample of the target audience for \name{}. The hypotheses are as follows:

\textbf{H1: Separate Design from Content.} The tool allows automatically handling the design component of infographics generation, with designers only having to control the infographics' content. 

\textbf{H2: Overall Layout Design.} The tool supports authoring and exploring multiple layouts, which can be used to design infographics from the given user content.

\textbf{H3: Visual Group Design.} The tool supports manipulating infographic designs on a VG level. Designers should be able to explore various VG designs and also customize infographics based on a user-provided VG design. 

\textbf{H4: Connection Design.} Like H3, the tool allows generating and exploring different connection designs and styles in infographic designs. The designers should optionally be able to remove any connections if required. 

\textbf{H5: Recommendations.} Following up on H1, the tool supports ranking different infographic designs given the content and design inputs. These rankings should provide an exploration of relevant infographic designs based on elements extracted from existing datasets.  

To the best of our knowledge, comparing existing tools according to table \ref{tab:comp_hyp} shows that \name{} satisfies all of the proposed infographic design hypotheses, unlike any other existing tools. Using our framework, \name{} supports flexible infographic designing with user input that combines the benefits of fully- and semi-automated features.

\subsection{Case Studies}
\begin{figure*}[tb]
    \centering
        \includegraphics[width=\textwidth]{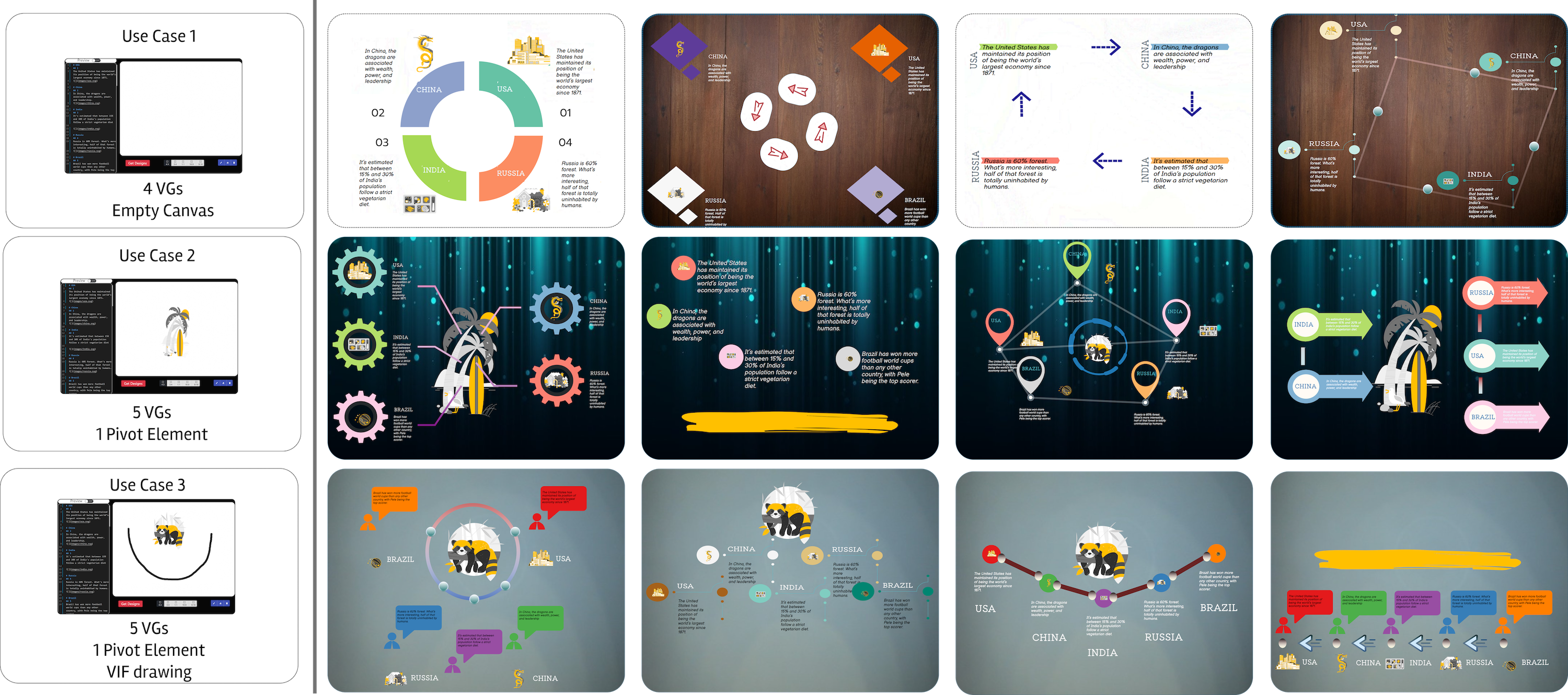}
    \caption{Infographics generated using our framework in the case studies. We evaluated our framework with three case studies, each of which had different content and design input, shown on the left of each row. For each input, our framework recommended infographics, where we show the top four recommended infographics in each row.} 
    \label{fig:case_study}
\end{figure*}

We demonstrate our framework's design capabilities using three use cases, covering the types of content and design feedback that can be an input to \name{} for generating infographics. As shown in Figure \ref{fig:case_study}, the three inputs to \name{} include: (a) markdown content with no design input on the canvas, (b) markdown content with a pivot element placed on the canvas, and (c) markdown content with a pivot element and a hand drawing for the VIF layout. Each row in Figure \ref{fig:case_study} shows the recommendations generated by \name{} corresponding to each input. High-resolution images can be found in our supplementary materials.

Further discussing the recommendations for each use case, for Use Case 1, the content includes information for four VGs with no design input on the canvas. In this case, \name{} is able to generate free designs varying the best-ranked VG designs, VIF layouts, and connection designs. In the example illustrations in Figure \ref{fig:case_study}, we show two background colors for this use case, white and brown, where \name{} provides contrasting color pallets for different background designs. 

For Use Case 2, the markdown content includes five VGs and a pivot graphic design input on the canvas. The recommendations, in this case, include a fixed pivot graphic, whereas the VIF layout, VG designs, and connection are ranked based on the content and the pivot element position. Figure \ref{fig:case_study} shows the generated recommendations with three different pivot elements for Use Case 2, where \name{} is able to generate infographic designs based on the shape of the pivot element. 

Finally, for Use Case 3, there is an additional design input of a VIF layout hand drawing along with the same input from Use Case 2. In this case, \name{} generates infographics by ranking the VIF layout, VG designs, and connection designs adhering to a fixed pivot graphic and VIF layouts similar to the hand drawing. As we can see from the generated recommendations in Figure \ref{fig:case_study}, the infographic designs follow the VIF layout similar to the hand drawing with all other design components varying in each infographic. 

Also, our tool generates designs by perfectly placing the markdown content inside varying VG designs with proper scaling and wrapping for all the use cases. Using \name{}, the designer can optionally change the background of the canvas, connection designs, and types, and VG color pallets using the menu bar on \name{} (Figure \ref{fig:teaser}). Following up on the results from all the use cases, we can conclude that our framework is able to generate infographic designs for varying design scenarios and use cases.

\subsection{User Study}
We further evaluated \name{} with real users for its ability to support the infographics design task. This study also aimed at investigating our framework's support for multiple factors of creativity. 
We did not conduct a comparative controlled study because we could not find comparable baselines that are publicly available to deploy and use. For commercial software, manual tools such as Adobe Illustrator is out of the scope of this work. Probably the most promising baseline is Microsoft PowerPoint, but it lacks the flexibility to control various design components (as indicated in Table \ref{tab:comp_hyp}), which is not a fair comparison. Thus, instead, this study's primary goal was to explore the strengths and weaknesses of \name{}, complementing the ACH comparison described in Section \ref{ss:ach}.

\subsubsection{Participants}
We recruited 10 participants (five males and five females, aged between 25 and 35 years) via social media and mailing lists. The participants were carefully chosen to be designers with different expertise levels. Of all the participants, five have over three years of designing experience with various mainstream tools like Adobe Photoshop and Adobe Illustrator, categorized as \textit{experts} for this study; the other five have zero to less than a year's experience in designing infographics, categorized as \textit{non-experts} for this study. 
Out of the total of five experts, two were Ph.D. students working in Data Visualization, and three work in the design industry. All the five non-experts were graduate students chosen based on their designing experience.

\subsubsection{Task and Procedure}
\label{ss:task1}

We initially familiarized the participants with the concepts of Infographics and related terminologies, such as the definition of VG, VIF, Pivot Graphic, and Connections for our framework. Next, we showed a few examples of infographics chosen from our dataset to familiarize them with infographic designs. The participants were then allowed to experiment with \name{} and ask clarifying questions regarding the tool. 
The task was to generate an infographic design from the content and design input of choice and evaluate the design support features of \name{}. We also provided sample markdown inputs to users, and they had an option to either generate infographics from the content of their choice or use the provided examples.  

After the participants were satisfied with their generated or filtered infographic designs, we conducted a short semi-structured interview to collect qualitative feedback. 
We asked the participants to rank their experience with \name{} on a five-point Likert scale. The questionnaire was based on a total of six factors: \textit{Enjoyment}, \textit{Exploration}, \textit{Expressiveness}, \textit{Results worth the effort}, \textit{Ease of use}, and \textit{Workflow}. Four of these factors were taken from the work by Cherry \textit{et al.} \cite{cherry2014quantifying} for quantifying the creativity support for design tools. Two specific factors, \textit{Effort} and \textit{Workflow}, were added in the questionnaire to specifically evaluate \name{} for its support in infographics design and exploration tasks. 
The whole study took about 45 minutes for each participant.

\subsubsection{Questionnaire Results}

\begin{figure}[tb]
    \centering
        \includegraphics[width=\textwidth]{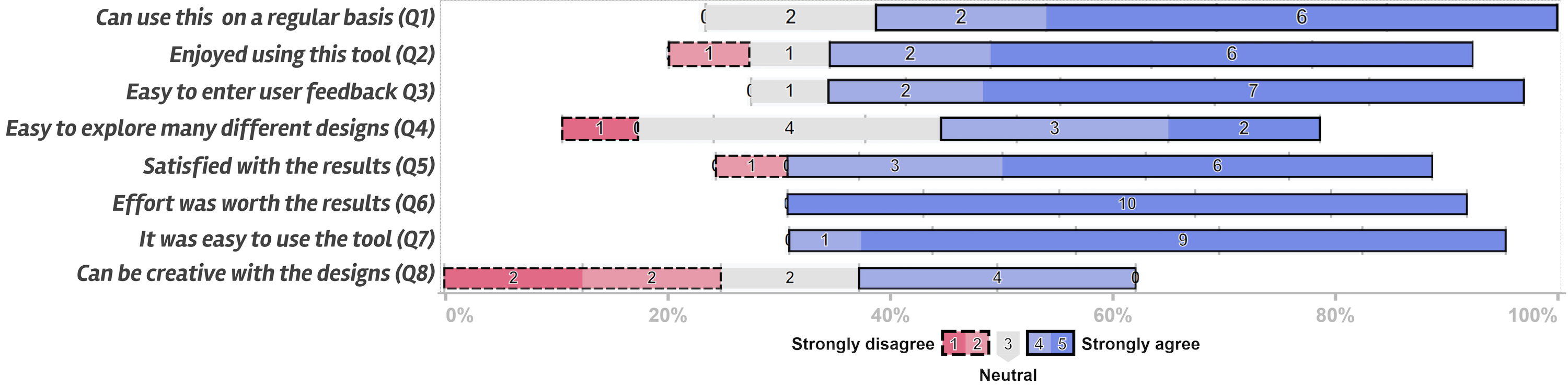}
    \caption{The user study results to evaluate our framework with the Creative Support Index \cite{cherry2014quantifying}, on a five-point Likert scale. We asked eight questions to the participants, covering different aspects of usability.} 
    \label{fig:likert}
\end{figure}

As shown in Figure \ref{fig:likert}, on average, eight out of 10 participants rated \name{} with a score of 4 and 5 for every question. Overall, our tool received 100 percent \textit{Agree votes} (scores of 4 and 5) on Q6 and Q7. Similarly, the second highest rated questions were Q1 and Q3. 

\begin{figure}[tb]
    \centering
        \includegraphics[width=\textwidth]{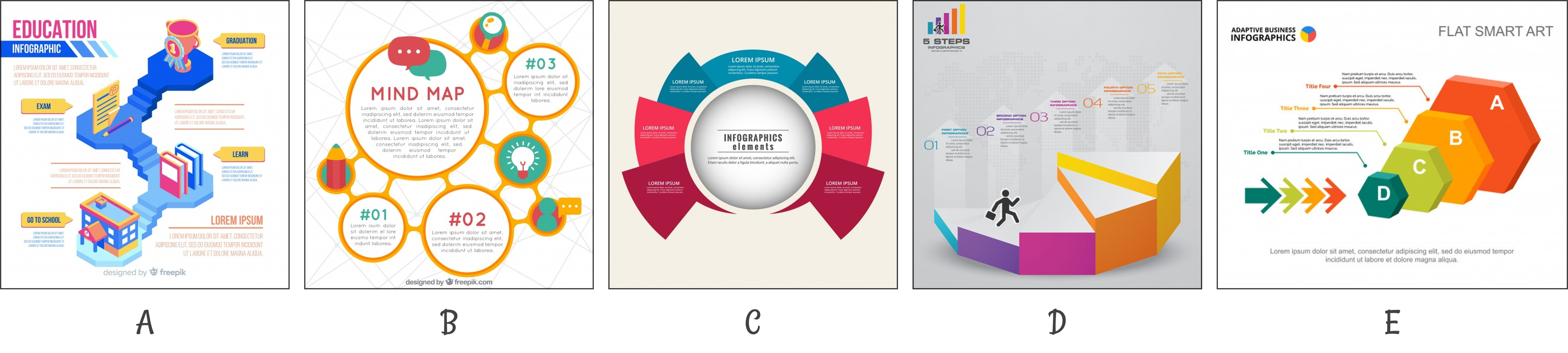}
    \caption{Infographic designs out of the scope of our framework. Some of these features include VG scaling (A, B, and C), inconsistent VG designs (A and B), complex pivot graphics composed of multiple SVG paths (A and D), and complex connection designs (E).} 
    \label{fig:fail_cases}
\end{figure}

Besides the general results, we separately collected feedback for the low-scoring (scores of 1 and 2) features from particular participants. 
Q8 was the weakest scoring question from this study because four of the experts were trying to generate highly complex infographics, which required using features not yet supported in \name{}. For example, in the infographics shown in Figure \ref{fig:fail_cases}, the VGs are scaled following the VIF layout in images \textit{A}, \textit{B}, and \textit{C}. Also, in images \textit{A} and \textit{C}, the pivot graphics are composed of multiple objects and contain additional icons. Sometimes, the VG design is not fixed, as in the case of image \textit{B}. Such designs are hard to generalize with our tool, which was developed to support general-purpose infographic design and exploration. Some sample infographics generated during the user study sessions are shown in Figure \ref{fig:collage}.


\subsubsection{Interview with Experts}
To further validate our framework for its design capabilities from an expert's eyes, we interviewed two expert designers (\textit{E1} and \textit{E2}) from the user study to collect in-depth qualitative feedback about our framework. The interview also helped compare \name{} with existing design tools as both the participants had more than five years of experience designing infographics using the existing design tools like Adobe Photoshop and Figma. We discuss our results organized by the themes of the questionnaire questions in the following.

\textbf{Ease of usage (Q7).} Both the participants found our framework easy to use for designing infographics. E1 explained \textit{``I really liked the simplistic design of the tool with a filtering and exploration panel on the side. It is effortless to navigate back and forth from design to exploration, similar to PowerPoint Design Ideas.''} E2 added that \textit{``I like the interface and the usage flow from left to right, moving from the content section on the left to the canvas and then finally the exploration section on the right.''}

\textbf{Usability (Q1, Q2, and Q6).} Both the experts found that the tool is handy in exploring different infographic designs. E1 commented that \textit{``I love the idea of separating the infographic content from design. The tool is really fast in generating infographic design recommendations for a given content.''} Similarly, E2 commented that \textit{``I can use \name{} as an exploration tool before designing very complex infographics to get new ideas. The fact that we can export final infographic designs as SVGs makes this tool very useful for designers.''}

\textbf{Quality of infographics (Q4, Q5 and Q8).} The participants were delighted with the variety of recommendations and the designs generated by our tool. E1 commented \textit{``The variety of VGs and Layouts is handy for exploration. I don't have to worry about finding the right template even for many VGs, which is a big problem in existing datasets. There are only a limited number of templates available for cases when the number of VGs exceeds 7.''} 
On similar lines, E2 commented about the designs generated for very complex layouts, saying \textit{``Now I can use the existing design of VGs to extend to very complex layouts. This was hard to achieve with existing infographic design tools as sometimes, we need an infographic design with a very unique layout. Most of the time, exploring infographic designs in such complex cases is impossible, and the designers have no ground truth to compare their work.''} E2 also commented about the creativity of the generated infographics saying, \textit{``I would have probably generated better infographics using Adobe Illustrator in some cases.''} However, they agreed that the amount of effort and time required to generate infographics would have increased. Instead, both participants suggested that they can export the designs and later edit with Adobe Illustrator to generate rare customized designs. 

\textbf{User feedback to the interface (Q3).} Both the participants were happy with the results generated after experimenting with our tool's design feedback feature. E1 commented about the pivot graphic functionality, saying, \textit{``I was surprised by the accuracy and speed of the final designs changing almost instantly as the pivot element is updated. It will be interesting to see the designs when this framework supports multiple pivot elements in the future.''} E2 commented, \textit{``I haven't seen any automatic infographic design tools where we can input so many design constraints. The ability to control the VG, VIF flow, connection designs, and even the color pallets gives the designers immense control over the design task. The tool is excellent in using the existing design pieces from large infographics datasets to generate very unique designs.''} However, E2 also mentioned that \textit{``it would be useful if the interface could support the designing of VGs as well, for custom input. Also, sometimes the connections and VGs are shifted by an offset in the generated infographics; in such cases, it would be useful to make minor changes in the generated infographic with the interface.''}

\section{Discussion, Conclusion, and Future Work}
\label{s:conclusion}
In this paper, we have presented an infographics authoring and exploration framework for a given user-input content. Our framework flexibly supports both fully automated infographic design with no design input and semi-automated infographic generation with design feedback, supporting designers with different levels of experiences. 

Overcoming the limitations of the existing design tools, our framework provides a comprehensive and sustainable infographic generation solution, realized in our tool, \name{}. Since we break down the infographic generation task into three independent pipeline steps, improving any one of these pipeline stages impacts the quality of the generated infographics directly. For example, we can only improve the VIF layout generation algorithm in the future without worrying about VG design and connection design rankings while still improving the generated infographic designs. Further, designers can also generate and explore infographics for very rare VG or VIF layout designs, which are not commonly found in the existing infographics templates. Our framework separates the content from the design aspect of infographic generation and reuses the design components from existing datasets to create new designs based on the user content. Hence, designers can now generate infographics for very complex layouts and even custom VG designs with minimal effort. We created an interface, \name{} to implement our framework, which can be used for design exploration. The infographic recommendations can be exported as SVG files for further low-level editing and fine-tuning with existing design tools like Adobe Illustrator. This feature is crucial in reducing the design time to generate very complex infographic designs.  

Besides the effectiveness of our framework, there remain limitations in some aspects regarding infographic design generation. Even though not widespread, there is a class of infographics that do not follow the general norm of placing information inside similar VGs, which cannot be designed with our tool. Factors like scaling of VGs inside an infographic, using multiple VG designs, complex multi-object pivot elements, and complex connection designs are not yet supported in our framework. From the design aspect, \name{} currently only supports a single pivot element. Also, for the VIF layout hand drawing feature of the canvas, the highly ranked matching VIF layouts, in some cases, do not match the users drawing direction since we do not use the drawing direction as an input to rank closely related VIF flows. These failure cases are discussed in more detail in the supplementary material. 

Following up on the limitations, we have included these tasks as a part of our future work. We also plan to expand our VG and VIF datasets by adding more infographics from different sources since this will directly impact the design quality of the infographics generated by our framework. Another part of our future research is to generate more generic infographics not following the norm of a single VG and have multiple coherent VG designs. 
Moreover, we wish to conduct more user evaluations to investigate our approach's advantages and disadvantages thoroughly. We plan to deploy \name{} with real users in a longer-term study to collect in-depth feedback and usage scenarios.

\bibliographystyle{ACM-Reference-Format}
\bibliography{paper_files/references}


\begin{thebibliography}{70}


\ifx \showCODEN    \undefined \def \showCODEN     #1{\unskip}     \fi
\ifx \showDOI      \undefined \def \showDOI       #1{#1}\fi
\ifx \showISBNx    \undefined \def \showISBNx     #1{\unskip}     \fi
\ifx \showISBNxiii \undefined \def \showISBNxiii  #1{\unskip}     \fi
\ifx \showISSN     \undefined \def \showISSN      #1{\unskip}     \fi
\ifx \showLCCN     \undefined \def \showLCCN      #1{\unskip}     \fi
\ifx \shownote     \undefined \def \shownote      #1{#1}          \fi
\ifx \showarticletitle \undefined \def \showarticletitle #1{#1}   \fi
\ifx \showURL      \undefined \def \showURL       {\relax}        \fi
\providecommand\bibfield[2]{#2}
\providecommand\bibinfo[2]{#2}
\providecommand\natexlab[1]{#1}
\providecommand\showeprint[2][]{arXiv:#2}

\bibitem[\protect\citeauthoryear{Adobe}{Adobe}{2021a}]%
        {photoshop}
\bibfield{author}{\bibinfo{person}{Adobe}.} \bibinfo{year}{2021}\natexlab{a}.
\newblock \bibinfo{title}{Adobe Photoshop}.
\newblock
\newblock
\urldef\tempurl%
\url{https://www.adobe.com/products/photoshopfamily.html}
\showURL{%
\tempurl}


\bibitem[\protect\citeauthoryear{Adobe}{Adobe}{2021b}]%
        {adobe}
\bibfield{author}{\bibinfo{person}{Adobe}.} \bibinfo{year}{2021}\natexlab{b}.
\newblock \bibinfo{title}{Adobe Stock}.
\newblock
\newblock
\urldef\tempurl%
\url{https://stock.adobe.com/}
\showURL{%
\tempurl}


\bibitem[\protect\citeauthoryear{Adobe}{Adobe}{2021c}]%
        {adobexd}
\bibfield{author}{\bibinfo{person}{Adobe}.} \bibinfo{year}{2021}\natexlab{c}.
\newblock \bibinfo{title}{Adobe XD}.
\newblock
\newblock
\urldef\tempurl%
\url{https://www.adobe.com/products/xd.html}
\showURL{%
\tempurl}


\bibitem[\protect\citeauthoryear{Amazon}{Amazon}{2021}]%
        {mturk}
\bibfield{author}{\bibinfo{person}{Amazon}.} \bibinfo{year}{2021}\natexlab{}.
\newblock \bibinfo{title}{Amazon Mechanical Turk}.
\newblock
\newblock
\urldef\tempurl%
\url{https://www.mturk.com/}
\showURL{%
\tempurl}


\bibitem[\protect\citeauthoryear{Amini, Riche, Lee, Monroy-Hernandez, and
  Irani}{Amini et~al\mbox{.}}{2016}]%
        {amini2016authoring}
\bibfield{author}{\bibinfo{person}{Fereshteh Amini},
  \bibinfo{person}{Nathalie~Henry Riche}, \bibinfo{person}{Bongshin Lee},
  \bibinfo{person}{Andres Monroy-Hernandez}, {and} \bibinfo{person}{Pourang
  Irani}.} \bibinfo{year}{2016}\natexlab{}.
\newblock \showarticletitle{Authoring data-driven videos with dataclips}.
\newblock \bibinfo{journal}{\emph{IEEE transactions on visualization and
  computer graphics}} \bibinfo{volume}{23}, \bibinfo{number}{1}
  (\bibinfo{year}{2016}), \bibinfo{pages}{501--510}.
\newblock


\bibitem[\protect\citeauthoryear{Bach, Kerracher, Hall, Carpendale, Kennedy,
  and Henry~Riche}{Bach et~al\mbox{.}}{2016}]%
        {bach2016telling}
\bibfield{author}{\bibinfo{person}{Benjamin Bach}, \bibinfo{person}{Natalie
  Kerracher}, \bibinfo{person}{Kyle~Wm Hall}, \bibinfo{person}{Sheelagh
  Carpendale}, \bibinfo{person}{Jessie Kennedy}, {and}
  \bibinfo{person}{Nathalie Henry~Riche}.} \bibinfo{year}{2016}\natexlab{}.
\newblock \showarticletitle{Telling stories about dynamic networks with graph
  comics}. In \bibinfo{booktitle}{\emph{Proceedings of the 2016 CHI Conference
  on Human Factors in Computing Systems}}. \bibinfo{pages}{3670--3682}.
\newblock


\bibitem[\protect\citeauthoryear{Bateman, Mandryk, Gutwin, Genest, McDine, and
  Brooks}{Bateman et~al\mbox{.}}{2010}]%
        {bateman2010useful}
\bibfield{author}{\bibinfo{person}{Scott Bateman}, \bibinfo{person}{Regan~L
  Mandryk}, \bibinfo{person}{Carl Gutwin}, \bibinfo{person}{Aaron Genest},
  \bibinfo{person}{David McDine}, {and} \bibinfo{person}{Christopher Brooks}.}
  \bibinfo{year}{2010}\natexlab{}.
\newblock \showarticletitle{Useful junk? The effects of visual embellishment on
  comprehension and memorability of charts}. In
  \bibinfo{booktitle}{\emph{Proceedings of the SIGCHI conference on human
  factors in computing systems}}. \bibinfo{pages}{2573--2582}.
\newblock


\bibitem[\protect\citeauthoryear{{Bongshin Lee}, Kazi, and Smith}{{Bongshin
  Lee} et~al\mbox{.}}{2013}]%
        {BongshinLee2013SketchStory}
\bibfield{author}{\bibinfo{person}{{Bongshin Lee}},
  \bibinfo{person}{Rubaiat~Habib Kazi}, {and} \bibinfo{person}{Greg Smith}.}
  \bibinfo{year}{2013}\natexlab{}.
\newblock \showarticletitle{{SketchStory}: {Telling} {More} {Engaging}
  {Stories} with {Data} through {Freeform} {Sketching}}.
\newblock \bibinfo{journal}{\emph{IEEE Transactions on Visualization and
  Computer Graphics}} \bibinfo{volume}{19}, \bibinfo{number}{12}
  (\bibinfo{date}{Dec.} \bibinfo{year}{2013}), \bibinfo{pages}{2416--2425}.
\newblock
\showISSN{1077-2626}
\urldef\tempurl%
\url{https://doi.org/10.1109/TVCG.2013.191}
\showDOI{\tempurl}


\bibitem[\protect\citeauthoryear{Borkin, Bylinskii, Kim, Bainbridge, Yeh,
  Borkin, Pfister, and Oliva}{Borkin et~al\mbox{.}}{2015}]%
        {borkin2015beyond}
\bibfield{author}{\bibinfo{person}{Michelle~A Borkin}, \bibinfo{person}{Zoya
  Bylinskii}, \bibinfo{person}{Nam~Wook Kim}, \bibinfo{person}{Constance~May
  Bainbridge}, \bibinfo{person}{Chelsea~S Yeh}, \bibinfo{person}{Daniel
  Borkin}, \bibinfo{person}{Hanspeter Pfister}, {and} \bibinfo{person}{Aude
  Oliva}.} \bibinfo{year}{2015}\natexlab{}.
\newblock \showarticletitle{Beyond memorability: Visualization recognition and
  recall}.
\newblock \bibinfo{journal}{\emph{IEEE transactions on visualization and
  computer graphics}} \bibinfo{volume}{22}, \bibinfo{number}{1}
  (\bibinfo{year}{2015}), \bibinfo{pages}{519--528}.
\newblock


\bibitem[\protect\citeauthoryear{Borkin, Vo, Bylinskii, Isola, Sunkavalli,
  Oliva, and Pfister}{Borkin et~al\mbox{.}}{2013}]%
        {borkin2013makes}
\bibfield{author}{\bibinfo{person}{Michelle~A Borkin},
  \bibinfo{person}{Azalea~A Vo}, \bibinfo{person}{Zoya Bylinskii},
  \bibinfo{person}{Phillip Isola}, \bibinfo{person}{Shashank Sunkavalli},
  \bibinfo{person}{Aude Oliva}, {and} \bibinfo{person}{Hanspeter Pfister}.}
  \bibinfo{year}{2013}\natexlab{}.
\newblock \showarticletitle{What makes a visualization memorable?}
\newblock \bibinfo{journal}{\emph{IEEE transactions on visualization and
  computer graphics}} \bibinfo{volume}{19}, \bibinfo{number}{12}
  (\bibinfo{year}{2013}), \bibinfo{pages}{2306--2315}.
\newblock


\bibitem[\protect\citeauthoryear{Brehmer, Lee, Riche, Tittsworth, Lytvynets,
  Edge, and White}{Brehmer et~al\mbox{.}}{2019}]%
        {brehmer2019timeline}
\bibfield{author}{\bibinfo{person}{Matthew Brehmer}, \bibinfo{person}{Bongshin
  Lee}, \bibinfo{person}{Nathalie~Henry Riche}, \bibinfo{person}{David
  Tittsworth}, \bibinfo{person}{Kate Lytvynets}, \bibinfo{person}{Darren Edge},
  {and} \bibinfo{person}{Christopher White}.} \bibinfo{year}{2019}\natexlab{}.
\newblock \showarticletitle{Timeline Storyteller}.
\newblock  (\bibinfo{year}{2019}).
\newblock


\bibitem[\protect\citeauthoryear{Bylinskii, Alsheikh, Madan, Recasens, Zhong,
  Pfister, Durand, and Oliva}{Bylinskii et~al\mbox{.}}{2017}]%
        {bylinskii2017understanding}
\bibfield{author}{\bibinfo{person}{Zoya Bylinskii}, \bibinfo{person}{Sami
  Alsheikh}, \bibinfo{person}{Spandan Madan}, \bibinfo{person}{Adria Recasens},
  \bibinfo{person}{Kimberli Zhong}, \bibinfo{person}{Hanspeter Pfister},
  \bibinfo{person}{Fredo Durand}, {and} \bibinfo{person}{Aude Oliva}.}
  \bibinfo{year}{2017}\natexlab{}.
\newblock \showarticletitle{Understanding infographics through textual and
  visual tag prediction}.
\newblock \bibinfo{journal}{\emph{arXiv preprint arXiv:1709.09215}}
  (\bibinfo{year}{2017}).
\newblock


\bibitem[\protect\citeauthoryear{Chen, Wang, Wang, Wang, and Qu}{Chen
  et~al\mbox{.}}{2019}]%
        {chen2019towards}
\bibfield{author}{\bibinfo{person}{Zhutian Chen}, \bibinfo{person}{Yun Wang},
  \bibinfo{person}{Qianwen Wang}, \bibinfo{person}{Yong Wang}, {and}
  \bibinfo{person}{Huamin Qu}.} \bibinfo{year}{2019}\natexlab{}.
\newblock \showarticletitle{Towards automated infographic design: Deep
  learning-based auto-extraction of extensible timeline}.
\newblock \bibinfo{journal}{\emph{IEEE transactions on visualization and
  computer graphics}} \bibinfo{volume}{26}, \bibinfo{number}{1}
  (\bibinfo{year}{2019}), \bibinfo{pages}{917--926}.
\newblock


\bibitem[\protect\citeauthoryear{Cherry and Latulipe}{Cherry and
  Latulipe}{2014}]%
        {cherry2014quantifying}
\bibfield{author}{\bibinfo{person}{Erin Cherry} {and} \bibinfo{person}{Celine
  Latulipe}.} \bibinfo{year}{2014}\natexlab{}.
\newblock \showarticletitle{Quantifying the creativity support of digital tools
  through the creativity support index}.
\newblock \bibinfo{journal}{\emph{ACM Transactions on Computer-Human
  Interaction (TOCHI)}} \bibinfo{volume}{21}, \bibinfo{number}{4}
  (\bibinfo{year}{2014}), \bibinfo{pages}{1--25}.
\newblock


\bibitem[\protect\citeauthoryear{Coelho and Mueller}{Coelho and
  Mueller}{2020}]%
        {coelho2020infomages}
\bibfield{author}{\bibinfo{person}{Darius Coelho} {and} \bibinfo{person}{Klaus
  Mueller}.} \bibinfo{year}{2020}\natexlab{}.
\newblock \showarticletitle{Infomages: Embedding Data into Thematic Images}. In
  \bibinfo{booktitle}{\emph{Computer Graphics Forum}},
  Vol.~\bibinfo{volume}{39}. Wiley Online Library, \bibinfo{pages}{593--606}.
\newblock


\bibitem[\protect\citeauthoryear{Cui, Zhang, Wang, Huang, Chen, Fang, Zhang,
  Lou, and Zhang}{Cui et~al\mbox{.}}{2019}]%
        {cui2019text}
\bibfield{author}{\bibinfo{person}{Weiwei Cui}, \bibinfo{person}{Xiaoyu Zhang},
  \bibinfo{person}{Yun Wang}, \bibinfo{person}{He Huang}, \bibinfo{person}{Bei
  Chen}, \bibinfo{person}{Lei Fang}, \bibinfo{person}{Haidong Zhang},
  \bibinfo{person}{Jian-Guan Lou}, {and} \bibinfo{person}{Dongmei Zhang}.}
  \bibinfo{year}{2019}\natexlab{}.
\newblock \showarticletitle{Text-to-viz: Automatic generation of infographics
  from proportion-related natural language statements}.
\newblock \bibinfo{journal}{\emph{IEEE transactions on visualization and
  computer graphics}} \bibinfo{volume}{26}, \bibinfo{number}{1}
  (\bibinfo{year}{2019}), \bibinfo{pages}{906--916}.
\newblock


\bibitem[\protect\citeauthoryear{Dibia and Demiralp}{Dibia and
  Demiralp}{2019}]%
        {dibia2019data2vis}
\bibfield{author}{\bibinfo{person}{Victor Dibia} {and}
  \bibinfo{person}{{\c{C}}a{\u{g}}atay Demiralp}.}
  \bibinfo{year}{2019}\natexlab{}.
\newblock \showarticletitle{Data2vis: Automatic generation of data
  visualizations using sequence-to-sequence recurrent neural networks}.
\newblock \bibinfo{journal}{\emph{IEEE computer graphics and applications}}
  \bibinfo{volume}{39}, \bibinfo{number}{5} (\bibinfo{year}{2019}),
  \bibinfo{pages}{33--46}.
\newblock


\bibitem[\protect\citeauthoryear{Dick}{Dick}{2014}]%
        {dick2014interactive}
\bibfield{author}{\bibinfo{person}{Murray Dick}.}
  \bibinfo{year}{2014}\natexlab{}.
\newblock \showarticletitle{Interactive infographics and news values}.
\newblock \bibinfo{journal}{\emph{Digital Journalism}} \bibinfo{volume}{2},
  \bibinfo{number}{4} (\bibinfo{year}{2014}), \bibinfo{pages}{490--506}.
\newblock


\bibitem[\protect\citeauthoryear{Fosco, Casser, Bedi, O'Donovan, Hertzmann, and
  Bylinskii}{Fosco et~al\mbox{.}}{2020}]%
        {Fosco2020}
\bibfield{author}{\bibinfo{person}{Camilo Fosco}, \bibinfo{person}{Vincent
  Casser}, \bibinfo{person}{Amish~Kumar Bedi}, \bibinfo{person}{Peter
  O'Donovan}, \bibinfo{person}{Aaron Hertzmann}, {and} \bibinfo{person}{Zoya
  Bylinskii}.} \bibinfo{year}{2020}\natexlab{}.
\newblock \showarticletitle{Predicting Visual Importance Across Graphic Design
  Types}. In \bibinfo{booktitle}{\emph{Proceedings of the 33rd Annual ACM
  Symposium on User Interface Software and Technology}} (Virtual Event, USA)
  \emph{(\bibinfo{series}{UIST '20})}. \bibinfo{publisher}{Association for
  Computing Machinery}, \bibinfo{address}{New York, NY, USA},
  \bibinfo{pages}{249–260}.
\newblock
\showISBNx{9781450375146}
\urldef\tempurl%
\url{https://doi.org/10.1145/3379337.3415825}
\showDOI{\tempurl}


\bibitem[\protect\citeauthoryear{Fulda, Brehmer, and Munzner}{Fulda
  et~al\mbox{.}}{2015}]%
        {fulda2015timelinecurator}
\bibfield{author}{\bibinfo{person}{Johanna Fulda}, \bibinfo{person}{Matthew
  Brehmer}, {and} \bibinfo{person}{Tamara Munzner}.}
  \bibinfo{year}{2015}\natexlab{}.
\newblock \showarticletitle{TimeLineCurator: Interactive authoring of visual
  timelines from unstructured text}.
\newblock \bibinfo{journal}{\emph{IEEE transactions on visualization and
  computer graphics}} \bibinfo{volume}{22}, \bibinfo{number}{1}
  (\bibinfo{year}{2015}), \bibinfo{pages}{300--309}.
\newblock


\bibitem[\protect\citeauthoryear{Haroz, Kosara, and Franconeri}{Haroz
  et~al\mbox{.}}{2015}]%
        {haroz2015isotype}
\bibfield{author}{\bibinfo{person}{Steve Haroz}, \bibinfo{person}{Robert
  Kosara}, {and} \bibinfo{person}{Steven~L Franconeri}.}
  \bibinfo{year}{2015}\natexlab{}.
\newblock \showarticletitle{Isotype visualization: Working memory, performance,
  and engagement with pictographs}. In \bibinfo{booktitle}{\emph{Proceedings of
  the 33rd annual ACM conference on human factors in computing systems}}.
  \bibinfo{pages}{1191--1200}.
\newblock


\bibitem[\protect\citeauthoryear{Harrison, Reinecke, and Chang}{Harrison
  et~al\mbox{.}}{2015}]%
        {harrison2015infographic}
\bibfield{author}{\bibinfo{person}{Lane Harrison}, \bibinfo{person}{Katharina
  Reinecke}, {and} \bibinfo{person}{Remco Chang}.}
  \bibinfo{year}{2015}\natexlab{}.
\newblock \showarticletitle{Infographic aesthetics: Designing for the first
  impression}. In \bibinfo{booktitle}{\emph{Proceedings of the 33rd Annual ACM
  Conference on Human Factors in Computing Systems}}.
  \bibinfo{pages}{1187--1190}.
\newblock


\bibitem[\protect\citeauthoryear{He, Gkioxari, Doll{\'a}r, and Girshick}{He
  et~al\mbox{.}}{2017}]%
        {he2017mask}
\bibfield{author}{\bibinfo{person}{Kaiming He}, \bibinfo{person}{Georgia
  Gkioxari}, \bibinfo{person}{Piotr Doll{\'a}r}, {and} \bibinfo{person}{Ross
  Girshick}.} \bibinfo{year}{2017}\natexlab{}.
\newblock \showarticletitle{Mask r-cnn}. In
  \bibinfo{booktitle}{\emph{Proceedings of the IEEE international conference on
  computer vision}}. \bibinfo{pages}{2961--2969}.
\newblock


\bibitem[\protect\citeauthoryear{Heuer~Jr}{Heuer~Jr}{1999}]%
        {heuer1999analysis}
\bibfield{author}{\bibinfo{person}{Richards~J Heuer~Jr}.}
  \bibinfo{year}{1999}\natexlab{}.
\newblock \showarticletitle{Analysis of competing hypotheses}.
\newblock \bibinfo{journal}{\emph{Psychology of intelligence analysis}}
  (\bibinfo{year}{1999}), \bibinfo{pages}{95--110}.
\newblock


\bibitem[\protect\citeauthoryear{Hu, Bakker, Li, Kraska, and Hidalgo}{Hu
  et~al\mbox{.}}{2019}]%
        {hu2019vizml}
\bibfield{author}{\bibinfo{person}{Kevin Hu}, \bibinfo{person}{Michiel~A
  Bakker}, \bibinfo{person}{Stephen Li}, \bibinfo{person}{Tim Kraska}, {and}
  \bibinfo{person}{C{\'e}sar Hidalgo}.} \bibinfo{year}{2019}\natexlab{}.
\newblock \showarticletitle{Vizml: A machine learning approach to visualization
  recommendation}. In \bibinfo{booktitle}{\emph{Proceedings of the 2019 CHI
  Conference on Human Factors in Computing Systems}}. \bibinfo{pages}{1--12}.
\newblock


\bibitem[\protect\citeauthoryear{Hullman, Drucker, Riche, Lee, Fisher, and
  Adar}{Hullman et~al\mbox{.}}{2013}]%
        {hullman2013deeper}
\bibfield{author}{\bibinfo{person}{Jessica Hullman}, \bibinfo{person}{Steven
  Drucker}, \bibinfo{person}{Nathalie~Henry Riche}, \bibinfo{person}{Bongshin
  Lee}, \bibinfo{person}{Danyel Fisher}, {and} \bibinfo{person}{Eytan Adar}.}
  \bibinfo{year}{2013}\natexlab{}.
\newblock \showarticletitle{A deeper understanding of sequence in narrative
  visualization}.
\newblock \bibinfo{journal}{\emph{IEEE Transactions on visualization and
  computer graphics}} \bibinfo{volume}{19}, \bibinfo{number}{12}
  (\bibinfo{year}{2013}), \bibinfo{pages}{2406--2415}.
\newblock


\bibitem[\protect\citeauthoryear{Joliffe and Morgan}{Joliffe and
  Morgan}{1992}]%
        {joliffe1992principal}
\bibfield{author}{\bibinfo{person}{Ian~T Joliffe} {and} \bibinfo{person}{BJT
  Morgan}.} \bibinfo{year}{1992}\natexlab{}.
\newblock \showarticletitle{Principal component analysis and exploratory factor
  analysis}.
\newblock \bibinfo{journal}{\emph{Statistical methods in medical research}}
  \bibinfo{volume}{1}, \bibinfo{number}{1} (\bibinfo{year}{1992}),
  \bibinfo{pages}{69--95}.
\newblock


\bibitem[\protect\citeauthoryear{Kim, Henry~Riche, Bach, Xu, Brehmer, Hinckley,
  Pahud, Xia, McGuffin, and Pfister}{Kim et~al\mbox{.}}{2019}]%
        {Kim2019DataToon}
\bibfield{author}{\bibinfo{person}{Nam~Wook Kim}, \bibinfo{person}{Nathalie
  Henry~Riche}, \bibinfo{person}{Benjamin Bach}, \bibinfo{person}{Guanpeng Xu},
  \bibinfo{person}{Matthew Brehmer}, \bibinfo{person}{Ken Hinckley},
  \bibinfo{person}{Michel Pahud}, \bibinfo{person}{Haijun Xia},
  \bibinfo{person}{Michael~J. McGuffin}, {and} \bibinfo{person}{Hanspeter
  Pfister}.} \bibinfo{year}{2019}\natexlab{}.
\newblock \showarticletitle{{DataToon}: {Drawing} {Dynamic} {Network} {Comics}
  {With} {Pen} + {Touch} {Interaction}}. In
  \bibinfo{booktitle}{\emph{Proceedings of the 2019 {CHI} {Conference} on
  {Human} {Factors} in {Computing} {Systems}}} \emph{(\bibinfo{series}{{CHI}
  '19})}. \bibinfo{publisher}{ACM}, \bibinfo{address}{New York, NY, USA},
  \bibinfo{pages}{105:1--105:12}.
\newblock
\showISBNx{978-1-4503-5970-2}
\urldef\tempurl%
\url{https://doi.org/10.1145/3290605.3300335}
\showDOI{\tempurl}
\newblock
\shownote{event-place: Glasgow, Scotland Uk.}


\bibitem[\protect\citeauthoryear{Kim, Schweickart, Liu, Dontcheva, Li, Popovic,
  and Pfister}{Kim et~al\mbox{.}}{2016}]%
        {kim2016data}
\bibfield{author}{\bibinfo{person}{Nam~Wook Kim}, \bibinfo{person}{Eston
  Schweickart}, \bibinfo{person}{Zhicheng Liu}, \bibinfo{person}{Mira
  Dontcheva}, \bibinfo{person}{Wilmot Li}, \bibinfo{person}{Jovan Popovic},
  {and} \bibinfo{person}{Hanspeter Pfister}.} \bibinfo{year}{2016}\natexlab{}.
\newblock \showarticletitle{Data-driven guides: Supporting expressive design
  for information graphics}.
\newblock \bibinfo{journal}{\emph{IEEE transactions on visualization and
  computer graphics}} \bibinfo{volume}{23}, \bibinfo{number}{1}
  (\bibinfo{year}{2016}), \bibinfo{pages}{491--500}.
\newblock


\bibitem[\protect\citeauthoryear{Kindlmann and Scheidegger}{Kindlmann and
  Scheidegger}{2014}]%
        {kindlmann2014algebraic}
\bibfield{author}{\bibinfo{person}{Gordon Kindlmann} {and}
  \bibinfo{person}{Carlos Scheidegger}.} \bibinfo{year}{2014}\natexlab{}.
\newblock \showarticletitle{An algebraic process for visualization design}.
\newblock \bibinfo{journal}{\emph{IEEE transactions on visualization and
  computer graphics}} \bibinfo{volume}{20}, \bibinfo{number}{12}
  (\bibinfo{year}{2014}), \bibinfo{pages}{2181--2190}.
\newblock


\bibitem[\protect\citeauthoryear{Kosara and Mackinlay}{Kosara and
  Mackinlay}{2013}]%
        {kosara2013storytelling}
\bibfield{author}{\bibinfo{person}{Robert Kosara} {and} \bibinfo{person}{Jock
  Mackinlay}.} \bibinfo{year}{2013}\natexlab{}.
\newblock \showarticletitle{Storytelling: The next step for visualization}.
\newblock \bibinfo{journal}{\emph{Computer}} \bibinfo{volume}{46},
  \bibinfo{number}{5} (\bibinfo{year}{2013}), \bibinfo{pages}{44--50}.
\newblock


\bibitem[\protect\citeauthoryear{Krum}{Krum}{2013}]%
        {krum2013cool}
\bibfield{author}{\bibinfo{person}{Randy Krum}.}
  \bibinfo{year}{2013}\natexlab{}.
\newblock \bibinfo{booktitle}{\emph{Cool infographics: Effective communication
  with data visualization and design}}.
\newblock \bibinfo{publisher}{John Wiley \& Sons}.
\newblock


\bibitem[\protect\citeauthoryear{Kumar, Tyagi, Burch, Weiskopf, and
  Mueller}{Kumar et~al\mbox{.}}{2019}]%
        {kumar2019task}
\bibfield{author}{\bibinfo{person}{Ayush Kumar}, \bibinfo{person}{Anjul Tyagi},
  \bibinfo{person}{Michael Burch}, \bibinfo{person}{Daniel Weiskopf}, {and}
  \bibinfo{person}{Klaus Mueller}.} \bibinfo{year}{2019}\natexlab{}.
\newblock \showarticletitle{Task classification model for visual fixation,
  exploration, and search}. In \bibinfo{booktitle}{\emph{Proceedings of the
  11th ACM Symposium on Eye Tracking Research \& Applications}}.
  \bibinfo{pages}{1--4}.
\newblock


\bibitem[\protect\citeauthoryear{Lankow, Ritchie, and Crooks}{Lankow
  et~al\mbox{.}}{2012}]%
        {lankow2012infographics}
\bibfield{author}{\bibinfo{person}{Jason Lankow}, \bibinfo{person}{Josh
  Ritchie}, {and} \bibinfo{person}{Ross Crooks}.}
  \bibinfo{year}{2012}\natexlab{}.
\newblock \bibinfo{booktitle}{\emph{Infographics: The power of visual
  storytelling}}.
\newblock \bibinfo{publisher}{John Wiley \& Sons}.
\newblock


\bibitem[\protect\citeauthoryear{Liu}{Liu}{2006}]%
        {liu2006fast}
\bibfield{author}{\bibinfo{person}{Bing Liu}.} \bibinfo{year}{2006}\natexlab{}.
\newblock \showarticletitle{A fast density-based clustering algorithm for large
  databases}. In \bibinfo{booktitle}{\emph{2006 International Conference on
  Machine Learning and Cybernetics}}. IEEE, \bibinfo{pages}{996--1000}.
\newblock


\bibitem[\protect\citeauthoryear{Liu, Thompson, Wilson, Dontcheva, Delorey,
  Grigg, Kerr, and Stasko}{Liu et~al\mbox{.}}{2018}]%
        {liu2018data}
\bibfield{author}{\bibinfo{person}{Zhicheng Liu}, \bibinfo{person}{John
  Thompson}, \bibinfo{person}{Alan Wilson}, \bibinfo{person}{Mira Dontcheva},
  \bibinfo{person}{James Delorey}, \bibinfo{person}{Sam Grigg},
  \bibinfo{person}{Bernard Kerr}, {and} \bibinfo{person}{John Stasko}.}
  \bibinfo{year}{2018}\natexlab{}.
\newblock \showarticletitle{Data Illustrator: Augmenting vector design tools
  with lazy data binding for expressive visualization authoring}. In
  \bibinfo{booktitle}{\emph{Proceedings of the 2018 CHI Conference on Human
  Factors in Computing Systems}}. \bibinfo{pages}{1--13}.
\newblock


\bibitem[\protect\citeauthoryear{Lu, Wang, Lanir, Zhao, Pfister, Cohen-Or, and
  Huang}{Lu et~al\mbox{.}}{2020}]%
        {lu2020exploring}
\bibfield{author}{\bibinfo{person}{Min Lu}, \bibinfo{person}{Chufeng Wang},
  \bibinfo{person}{Joel Lanir}, \bibinfo{person}{Nanxuan Zhao},
  \bibinfo{person}{Hanspeter Pfister}, \bibinfo{person}{Daniel Cohen-Or}, {and}
  \bibinfo{person}{Hui Huang}.} \bibinfo{year}{2020}\natexlab{}.
\newblock \showarticletitle{Exploring visual information flows in
  infographics}. In \bibinfo{booktitle}{\emph{Proceedings of the 2020 CHI
  Conference on Human Factors in Computing Systems}}. \bibinfo{pages}{1--12}.
\newblock


\bibitem[\protect\citeauthoryear{Luo, Qin, Tang, and Li}{Luo
  et~al\mbox{.}}{2018}]%
        {luo2018deepeye}
\bibfield{author}{\bibinfo{person}{Yuyu Luo}, \bibinfo{person}{Xuedi Qin},
  \bibinfo{person}{Nan Tang}, {and} \bibinfo{person}{Guoliang Li}.}
  \bibinfo{year}{2018}\natexlab{}.
\newblock \showarticletitle{Deepeye: Towards automatic data visualization}. In
  \bibinfo{booktitle}{\emph{2018 IEEE 34th international conference on data
  engineering (ICDE)}}. IEEE, \bibinfo{pages}{101--112}.
\newblock


\bibitem[\protect\citeauthoryear{Mackinlay}{Mackinlay}{1986}]%
        {mackinlay1986automating}
\bibfield{author}{\bibinfo{person}{Jock Mackinlay}.}
  \bibinfo{year}{1986}\natexlab{}.
\newblock \showarticletitle{Automating the design of graphical presentations of
  relational information}.
\newblock \bibinfo{journal}{\emph{Acm Transactions On Graphics (Tog)}}
  \bibinfo{volume}{5}, \bibinfo{number}{2} (\bibinfo{year}{1986}),
  \bibinfo{pages}{110--141}.
\newblock


\bibitem[\protect\citeauthoryear{Madan, Bylinskii, Tancik, Recasens, Zhong,
  Alsheikh, Pfister, Oliva, and Durand}{Madan et~al\mbox{.}}{2018}]%
        {madan2018synthetically}
\bibfield{author}{\bibinfo{person}{Spandan Madan}, \bibinfo{person}{Zoya
  Bylinskii}, \bibinfo{person}{Matthew Tancik}, \bibinfo{person}{Adri{\`a}
  Recasens}, \bibinfo{person}{Kimberli Zhong}, \bibinfo{person}{Sami Alsheikh},
  \bibinfo{person}{Hanspeter Pfister}, \bibinfo{person}{Aude Oliva}, {and}
  \bibinfo{person}{Fredo Durand}.} \bibinfo{year}{2018}\natexlab{}.
\newblock \showarticletitle{Synthetically trained icon proposals for parsing
  and summarizing infographics}.
\newblock \bibinfo{journal}{\emph{arXiv preprint arXiv:1807.10441}}
  (\bibinfo{year}{2018}).
\newblock


\bibitem[\protect\citeauthoryear{Marvel}{Marvel}{2021}]%
        {marvel}
\bibfield{author}{\bibinfo{person}{Marvel}.} \bibinfo{year}{2021}\natexlab{}.
\newblock \bibinfo{title}{Marvel}.
\newblock
\newblock
\urldef\tempurl%
\url{https://marvelapp.com/}
\showURL{%
\tempurl}


\bibitem[\protect\citeauthoryear{Mockflow}{Mockflow}{2021}]%
        {mockflow}
\bibfield{author}{\bibinfo{person}{Mockflow}.} \bibinfo{year}{2021}\natexlab{}.
\newblock \bibinfo{title}{Mockflow Wireframe}.
\newblock
\newblock
\urldef\tempurl%
\url{https://www.mockflow.com/}
\showURL{%
\tempurl}


\bibitem[\protect\citeauthoryear{Moere and Purchase}{Moere and
  Purchase}{2011}]%
        {moere2011role}
\bibfield{author}{\bibinfo{person}{Andrew~Vande Moere} {and}
  \bibinfo{person}{Helen Purchase}.} \bibinfo{year}{2011}\natexlab{}.
\newblock \showarticletitle{On the role of design in information
  visualization}.
\newblock \bibinfo{journal}{\emph{Information Visualization}}
  \bibinfo{volume}{10}, \bibinfo{number}{4} (\bibinfo{year}{2011}),
  \bibinfo{pages}{356--371}.
\newblock


\bibitem[\protect\citeauthoryear{Moritz, Wang, Nelson, Lin, Smith, Howe, and
  Heer}{Moritz et~al\mbox{.}}{2018}]%
        {moritz2018formalizing}
\bibfield{author}{\bibinfo{person}{Dominik Moritz}, \bibinfo{person}{Chenglong
  Wang}, \bibinfo{person}{Greg~L Nelson}, \bibinfo{person}{Halden Lin},
  \bibinfo{person}{Adam~M Smith}, \bibinfo{person}{Bill Howe}, {and}
  \bibinfo{person}{Jeffrey Heer}.} \bibinfo{year}{2018}\natexlab{}.
\newblock \showarticletitle{Formalizing visualization design knowledge as
  constraints: Actionable and extensible models in draco}.
\newblock \bibinfo{journal}{\emph{IEEE transactions on visualization and
  computer graphics}} \bibinfo{volume}{25}, \bibinfo{number}{1}
  (\bibinfo{year}{2018}), \bibinfo{pages}{438--448}.
\newblock


\bibitem[\protect\citeauthoryear{Munzner}{Munzner}{2009}]%
        {munzner2009nested}
\bibfield{author}{\bibinfo{person}{Tamara Munzner}.}
  \bibinfo{year}{2009}\natexlab{}.
\newblock \showarticletitle{A nested model for visualization design and
  validation}.
\newblock \bibinfo{journal}{\emph{IEEE transactions on visualization and
  computer graphics}} \bibinfo{volume}{15}, \bibinfo{number}{6}
  (\bibinfo{year}{2009}), \bibinfo{pages}{921--928}.
\newblock


\bibitem[\protect\citeauthoryear{Niebaum, Cunningham-Sabo, Carroll, and
  Bellows}{Niebaum et~al\mbox{.}}{2015}]%
        {niebaum2015infographics}
\bibfield{author}{\bibinfo{person}{Kelly Niebaum}, \bibinfo{person}{Leslie
  Cunningham-Sabo}, \bibinfo{person}{Jan Carroll}, {and} \bibinfo{person}{Laura
  Bellows}.} \bibinfo{year}{2015}\natexlab{}.
\newblock \showarticletitle{Infographics: An Innovative Tool to Capture
  Consumers' Attention.}
\newblock \bibinfo{journal}{\emph{Journal of extension}} \bibinfo{volume}{53},
  \bibinfo{number}{6} (\bibinfo{year}{2015}), \bibinfo{pages}{n6}.
\newblock


\bibitem[\protect\citeauthoryear{O'Donovan, Agarwala, and Hertzmann}{O'Donovan
  et~al\mbox{.}}{2015}]%
        {o2015designscape}
\bibfield{author}{\bibinfo{person}{Peter O'Donovan}, \bibinfo{person}{Aseem
  Agarwala}, {and} \bibinfo{person}{Aaron Hertzmann}.}
  \bibinfo{year}{2015}\natexlab{}.
\newblock \showarticletitle{Designscape: Design with interactive layout
  suggestions}. In \bibinfo{booktitle}{\emph{Proceedings of the 33rd annual ACM
  conference on human factors in computing systems}}.
  \bibinfo{pages}{1221--1224}.
\newblock


\bibitem[\protect\citeauthoryear{Park, Kaufman, and Mueller}{Park
  et~al\mbox{.}}{2018}]%
        {park2018graphoto}
\bibfield{author}{\bibinfo{person}{Ji~Hwan Park}, \bibinfo{person}{Arie
  Kaufman}, {and} \bibinfo{person}{Klaus Mueller}.}
  \bibinfo{year}{2018}\natexlab{}.
\newblock \showarticletitle{Graphoto: Aesthetically pleasing charts for casual
  information visualization}.
\newblock \bibinfo{journal}{\emph{IEEE computer graphics and applications}}
  \bibinfo{volume}{38}, \bibinfo{number}{6} (\bibinfo{year}{2018}),
  \bibinfo{pages}{67--82}.
\newblock


\bibitem[\protect\citeauthoryear{Piktochart}{Piktochart}{2021}]%
        {piktochart}
\bibfield{author}{\bibinfo{person}{Piktochart}.}
  \bibinfo{year}{2021}\natexlab{}.
\newblock \bibinfo{title}{Piktochart}.
\newblock
\newblock
\urldef\tempurl%
\url{https://piktochart.com/}
\showURL{%
\tempurl}


\bibitem[\protect\citeauthoryear{Proto}{Proto}{2021}]%
        {proto}
\bibfield{author}{\bibinfo{person}{Proto}.} \bibinfo{year}{2021}\natexlab{}.
\newblock \bibinfo{title}{Proto.io}.
\newblock
\newblock
\urldef\tempurl%
\url{https://proto.io/}
\showURL{%
\tempurl}


\bibitem[\protect\citeauthoryear{Ramos et~al\mbox{.}}{Ramos
  et~al\mbox{.}}{2003}]%
        {ramos2003using}
\bibfield{author}{\bibinfo{person}{Juan Ramos} {et~al\mbox{.}}}
  \bibinfo{year}{2003}\natexlab{}.
\newblock \showarticletitle{Using tf-idf to determine word relevance in
  document queries}. In \bibinfo{booktitle}{\emph{Proceedings of the first
  instructional conference on machine learning}}, Vol.~\bibinfo{volume}{242}.
  Citeseer, \bibinfo{pages}{29--48}.
\newblock


\bibitem[\protect\citeauthoryear{Ren, Brehmer, {Bongshin Lee}, Hollerer, and
  Choe}{Ren et~al\mbox{.}}{2017}]%
        {Ren2017ChartAccent}
\bibfield{author}{\bibinfo{person}{Donghao Ren}, \bibinfo{person}{Matthew
  Brehmer}, \bibinfo{person}{{Bongshin Lee}}, \bibinfo{person}{Tobias
  Hollerer}, {and} \bibinfo{person}{Eun~Kyoung Choe}.}
  \bibinfo{year}{2017}\natexlab{}.
\newblock \showarticletitle{{ChartAccent}: {Annotation} for data-driven
  storytelling}. In \bibinfo{booktitle}{\emph{2017 {IEEE} {Pacific}
  {Visualization} {Symposium} ({PacificVis})}}. \bibinfo{publisher}{IEEE},
  \bibinfo{address}{Seoul, South Korea}, \bibinfo{pages}{230--239}.
\newblock
\showISBNx{978-1-5090-5738-2}
\urldef\tempurl%
\url{https://doi.org/10.1109/PACIFICVIS.2017.8031599}
\showDOI{\tempurl}


\bibitem[\protect\citeauthoryear{Roth, Kolojejchick, Mattis, and
  Goldstein}{Roth et~al\mbox{.}}{1994}]%
        {roth1994interactive}
\bibfield{author}{\bibinfo{person}{Steven~F Roth}, \bibinfo{person}{John
  Kolojejchick}, \bibinfo{person}{Joe Mattis}, {and} \bibinfo{person}{Jade
  Goldstein}.} \bibinfo{year}{1994}\natexlab{}.
\newblock \showarticletitle{Interactive graphic design using automatic
  presentation knowledge}. In \bibinfo{booktitle}{\emph{Proceedings of the
  SIGCHI conference on Human factors in computing systems}}.
  \bibinfo{pages}{112--117}.
\newblock


\bibitem[\protect\citeauthoryear{Rother, Kolmogorov, and Blake}{Rother
  et~al\mbox{.}}{2004}]%
        {rother2004grabcut}
\bibfield{author}{\bibinfo{person}{Carsten Rother}, \bibinfo{person}{Vladimir
  Kolmogorov}, {and} \bibinfo{person}{Andrew Blake}.}
  \bibinfo{year}{2004}\natexlab{}.
\newblock \showarticletitle{" GrabCut" interactive foreground extraction using
  iterated graph cuts}.
\newblock \bibinfo{journal}{\emph{ACM transactions on graphics (TOG)}}
  \bibinfo{volume}{23}, \bibinfo{number}{3} (\bibinfo{year}{2004}),
  \bibinfo{pages}{309--314}.
\newblock


\bibitem[\protect\citeauthoryear{Satyanarayan, Moritz, Wongsuphasawat, and
  Heer}{Satyanarayan et~al\mbox{.}}{2016}]%
        {satyanarayan2016vega}
\bibfield{author}{\bibinfo{person}{Arvind Satyanarayan},
  \bibinfo{person}{Dominik Moritz}, \bibinfo{person}{Kanit Wongsuphasawat},
  {and} \bibinfo{person}{Jeffrey Heer}.} \bibinfo{year}{2016}\natexlab{}.
\newblock \showarticletitle{Vega-lite: A grammar of interactive graphics}.
\newblock \bibinfo{journal}{\emph{IEEE transactions on visualization and
  computer graphics}} \bibinfo{volume}{23}, \bibinfo{number}{1}
  (\bibinfo{year}{2016}), \bibinfo{pages}{341--350}.
\newblock


\bibitem[\protect\citeauthoryear{Satyanarayan, Russell, Hoffswell, and
  Heer}{Satyanarayan et~al\mbox{.}}{2015}]%
        {satyanarayan2015reactive}
\bibfield{author}{\bibinfo{person}{Arvind Satyanarayan}, \bibinfo{person}{Ryan
  Russell}, \bibinfo{person}{Jane Hoffswell}, {and} \bibinfo{person}{Jeffrey
  Heer}.} \bibinfo{year}{2015}\natexlab{}.
\newblock \showarticletitle{Reactive vega: A streaming dataflow architecture
  for declarative interactive visualization}.
\newblock \bibinfo{journal}{\emph{IEEE transactions on visualization and
  computer graphics}} \bibinfo{volume}{22}, \bibinfo{number}{1}
  (\bibinfo{year}{2015}), \bibinfo{pages}{659--668}.
\newblock


\bibitem[\protect\citeauthoryear{Segel and Heer}{Segel and Heer}{2010}]%
        {segel2010narrative}
\bibfield{author}{\bibinfo{person}{Edward Segel} {and} \bibinfo{person}{Jeffrey
  Heer}.} \bibinfo{year}{2010}\natexlab{}.
\newblock \showarticletitle{Narrative visualization: Telling stories with
  data}.
\newblock \bibinfo{journal}{\emph{IEEE transactions on visualization and
  computer graphics}} \bibinfo{volume}{16}, \bibinfo{number}{6}
  (\bibinfo{year}{2010}), \bibinfo{pages}{1139--1148}.
\newblock


\bibitem[\protect\citeauthoryear{Skau and Kosara}{Skau and Kosara}{2017}]%
        {skau2017readability}
\bibfield{author}{\bibinfo{person}{Drew Skau} {and} \bibinfo{person}{Robert
  Kosara}.} \bibinfo{year}{2017}\natexlab{}.
\newblock \showarticletitle{Readability and precision in pictorial bar charts}.
  In \bibinfo{booktitle}{\emph{Proceedings of the Eurographics/IEEE VGTC
  Conference on Visualization: Short Papers}}. \bibinfo{pages}{91--95}.
\newblock


\bibitem[\protect\citeauthoryear{Solaris}{Solaris}{2021}]%
        {solaris}
\bibfield{author}{\bibinfo{person}{Solaris}.} \bibinfo{year}{2021}\natexlab{}.
\newblock \bibinfo{title}{Solaris by CosmiQ Works}.
\newblock
\newblock
\urldef\tempurl%
\url{https://solaris.readthedocs.io/en/latest/index.html}
\showURL{%
\tempurl}


\bibitem[\protect\citeauthoryear{Steele and Iliinsky}{Steele and
  Iliinsky}{2011}]%
        {steele2011designing}
\bibfield{author}{\bibinfo{person}{Julie Steele} {and} \bibinfo{person}{Noah
  Iliinsky}.} \bibinfo{year}{2011}\natexlab{}.
\newblock \bibinfo{booktitle}{\emph{Designing data visualizations: Representing
  informational relationships}}.
\newblock \bibinfo{publisher}{O'Reilly Media}.
\newblock


\bibitem[\protect\citeauthoryear{Teh and Chin}{Teh and Chin}{1989}]%
        {teh1989detection}
\bibfield{author}{\bibinfo{person}{C-H Teh} {and} \bibinfo{person}{Roland~T.
  Chin}.} \bibinfo{year}{1989}\natexlab{}.
\newblock \showarticletitle{On the detection of dominant points on digital
  curves}.
\newblock \bibinfo{journal}{\emph{IEEE Transactions on pattern analysis and
  machine intelligence}} \bibinfo{volume}{11}, \bibinfo{number}{8}
  (\bibinfo{year}{1989}), \bibinfo{pages}{859--872}.
\newblock


\bibitem[\protect\citeauthoryear{Van~der Maaten and Hinton}{Van~der Maaten and
  Hinton}{2008}]%
        {van2008visualizing}
\bibfield{author}{\bibinfo{person}{Laurens Van~der Maaten} {and}
  \bibinfo{person}{Geoffrey Hinton}.} \bibinfo{year}{2008}\natexlab{}.
\newblock \showarticletitle{Visualizing data using t-SNE.}
\newblock \bibinfo{journal}{\emph{Journal of machine learning research}}
  \bibinfo{volume}{9}, \bibinfo{number}{11} (\bibinfo{year}{2008}).
\newblock


\bibitem[\protect\citeauthoryear{Wang, Sun, Zhang, Cui, Xu, Ma, and Zhang}{Wang
  et~al\mbox{.}}{2019}]%
        {wang2019datashot}
\bibfield{author}{\bibinfo{person}{Yun Wang}, \bibinfo{person}{Zhida Sun},
  \bibinfo{person}{Haidong Zhang}, \bibinfo{person}{Weiwei Cui},
  \bibinfo{person}{Ke Xu}, \bibinfo{person}{Xiaojuan Ma}, {and}
  \bibinfo{person}{Dongmei Zhang}.} \bibinfo{year}{2019}\natexlab{}.
\newblock \showarticletitle{DataShot: Automatic generation of fact sheets from
  tabular data}.
\newblock \bibinfo{journal}{\emph{IEEE transactions on visualization and
  computer graphics}} \bibinfo{volume}{26}, \bibinfo{number}{1}
  (\bibinfo{year}{2019}), \bibinfo{pages}{895--905}.
\newblock


\bibitem[\protect\citeauthoryear{Wang, Zhang, Huang, Chen, Yin, Hou, Zhang,
  Luo, and Qu}{Wang et~al\mbox{.}}{2018}]%
        {wang2018infonice}
\bibfield{author}{\bibinfo{person}{Yun Wang}, \bibinfo{person}{Haidong Zhang},
  \bibinfo{person}{He Huang}, \bibinfo{person}{Xi Chen},
  \bibinfo{person}{Qiufeng Yin}, \bibinfo{person}{Zhitao Hou},
  \bibinfo{person}{Dongmei Zhang}, \bibinfo{person}{Qiong Luo}, {and}
  \bibinfo{person}{Huamin Qu}.} \bibinfo{year}{2018}\natexlab{}.
\newblock \showarticletitle{InfoNice: Easy creation of information graphics}.
  In \bibinfo{booktitle}{\emph{Proceedings of the 2018 CHI Conference on Human
  Factors in Computing Systems}}. \bibinfo{pages}{1--12}.
\newblock


\bibitem[\protect\citeauthoryear{Wongsuphasawat, Moritz, Anand, Mackinlay,
  Howe, and Heer}{Wongsuphasawat et~al\mbox{.}}{2015}]%
        {wongsuphasawat2015voyager}
\bibfield{author}{\bibinfo{person}{Kanit Wongsuphasawat},
  \bibinfo{person}{Dominik Moritz}, \bibinfo{person}{Anushka Anand},
  \bibinfo{person}{Jock Mackinlay}, \bibinfo{person}{Bill Howe}, {and}
  \bibinfo{person}{Jeffrey Heer}.} \bibinfo{year}{2015}\natexlab{}.
\newblock \showarticletitle{Voyager: Exploratory analysis via faceted browsing
  of visualization recommendations}.
\newblock \bibinfo{journal}{\emph{IEEE transactions on visualization and
  computer graphics}} \bibinfo{volume}{22}, \bibinfo{number}{1}
  (\bibinfo{year}{2015}), \bibinfo{pages}{649--658}.
\newblock


\bibitem[\protect\citeauthoryear{Wongsuphasawat, Moritz, Anand, Mackinlay,
  Howe, and Heer}{Wongsuphasawat et~al\mbox{.}}{2016}]%
        {wongsuphasawat2016towards}
\bibfield{author}{\bibinfo{person}{Kanit Wongsuphasawat},
  \bibinfo{person}{Dominik Moritz}, \bibinfo{person}{Anushka Anand},
  \bibinfo{person}{Jock Mackinlay}, \bibinfo{person}{Bill Howe}, {and}
  \bibinfo{person}{Jeffrey Heer}.} \bibinfo{year}{2016}\natexlab{}.
\newblock \showarticletitle{Towards a general-purpose query language for
  visualization recommendation}. In \bibinfo{booktitle}{\emph{Proceedings of
  the Workshop on Human-In-the-Loop Data Analytics}}. \bibinfo{pages}{1--6}.
\newblock


\bibitem[\protect\citeauthoryear{Wongsuphasawat, Qu, Moritz, Chang, Ouk, Anand,
  Mackinlay, Howe, and Heer}{Wongsuphasawat et~al\mbox{.}}{2017}]%
        {wongsuphasawat2017voyager}
\bibfield{author}{\bibinfo{person}{Kanit Wongsuphasawat},
  \bibinfo{person}{Zening Qu}, \bibinfo{person}{Dominik Moritz},
  \bibinfo{person}{Riley Chang}, \bibinfo{person}{Felix Ouk},
  \bibinfo{person}{Anushka Anand}, \bibinfo{person}{Jock Mackinlay},
  \bibinfo{person}{Bill Howe}, {and} \bibinfo{person}{Jeffrey Heer}.}
  \bibinfo{year}{2017}\natexlab{}.
\newblock \showarticletitle{Voyager 2: Augmenting visual analysis with partial
  view specifications}. In \bibinfo{booktitle}{\emph{Proceedings of the 2017
  CHI Conference on Human Factors in Computing Systems}}.
  \bibinfo{pages}{2648--2659}.
\newblock


\bibitem[\protect\citeauthoryear{Xia, Henry~Riche, Chevalier, De~Araujo, and
  Wigdor}{Xia et~al\mbox{.}}{2018}]%
        {xia2018dataink}
\bibfield{author}{\bibinfo{person}{Haijun Xia}, \bibinfo{person}{Nathalie
  Henry~Riche}, \bibinfo{person}{Fanny Chevalier}, \bibinfo{person}{Bruno
  De~Araujo}, {and} \bibinfo{person}{Daniel Wigdor}.}
  \bibinfo{year}{2018}\natexlab{}.
\newblock \showarticletitle{DataInk: Direct and creative data-oriented
  drawing}. In \bibinfo{booktitle}{\emph{Proceedings of the 2018 CHI Conference
  on Human Factors in Computing Systems}}. \bibinfo{pages}{1--13}.
\newblock


\bibitem[\protect\citeauthoryear{Zhang, Sultanum, Bezerianos, and
  Chevalier}{Zhang et~al\mbox{.}}{2020}]%
        {zhang2020dataquilta}
\bibfield{author}{\bibinfo{person}{Jiayi~Eris Zhang}, \bibinfo{person}{Nicole
  Sultanum}, \bibinfo{person}{Anastasia Bezerianos}, {and}
  \bibinfo{person}{Fanny Chevalier}.} \bibinfo{year}{2020}\natexlab{}.
\newblock \showarticletitle{{DataQuilt}: {Extracting} {Visual} {Elements} from
  {Images} to {Craft} {Pictorial} {Visualizations}}. In
  \bibinfo{booktitle}{\emph{Proceedings of the 2020 {CHI} {Conference} on
  {Human} {Factors} in {Computing} {Systems}}} \emph{(\bibinfo{series}{{CHI}
  '20})}. \bibinfo{publisher}{Association for Computing Machinery},
  \bibinfo{address}{Honolulu, HI, USA}, \bibinfo{pages}{1--13}.
\newblock
\showISBNx{978-1-4503-6708-0}
\urldef\tempurl%
\url{https://doi.org/10.1145/3313831.3376172}
\showDOI{\tempurl}


\bibitem[\protect\citeauthoryear{Zhao, Kim, Herman, Pfister, Lau, Echevarria,
  and Bylinskii}{Zhao et~al\mbox{.}}{2020}]%
        {zhao2020iconate}
\bibfield{author}{\bibinfo{person}{Nanxuan Zhao}, \bibinfo{person}{Nam~Wook
  Kim}, \bibinfo{person}{Laura~Mariah Herman}, \bibinfo{person}{Hanspeter
  Pfister}, \bibinfo{person}{Rynson~WH Lau}, \bibinfo{person}{Jose Echevarria},
  {and} \bibinfo{person}{Zoya Bylinskii}.} \bibinfo{year}{2020}\natexlab{}.
\newblock \showarticletitle{Iconate: Automatic compound icon generation and
  ideation}. In \bibinfo{booktitle}{\emph{Proceedings of the 2020 CHI
  Conference on Human Factors in Computing Systems}}. \bibinfo{pages}{1--13}.
\newblock


\end{thebibliography}

\end{document}